\documentclass[aps,prd,amsmath,amssymb]{revtex4}

\usepackage{epsfig,float}
\usepackage{graphicx}
\usepackage{subfigure}
\usepackage{dcolumn}
\usepackage{morefloats}
\usepackage{color}
\usepackage{slashed}
\usepackage{bbm}
\usepackage{bm}
\usepackage{mathtools}
\usepackage{booktabs}
\usepackage{multirow}
\usepackage{mathrsfs}
\usepackage{verbatim} 
\usepackage{textcomp,nicefrac}
\allowdisplaybreaks[4]

\newcommand{\la}{\left\langle}
\newcommand{\ra}{\right\rangle}

\newcommand{\be}{\begin{equation}}
\newcommand{\ee}{\end{equation}}
\newcommand{\ba}{\begin{eqnarray}}
\newcommand{\ea}{\end{eqnarray}}

\newcommand{\half}{\nicefrac{1}{2}}
\begin{document}

\title{ Hadronic weak decays of $\Lambda_c$ in the quark model}

\author{Peng-Yu Niu$^{1,2}$\footnote{Email: niupy@ihep.ac.cn}, Jean-Marc Richard$^3$\footnote{E-mail: j-m.richard@ipnl.in2p3.fr}, Qian Wang$^{4,1}$\footnote{E-mail: qianwang@m.scnu.edu.cn}, Qiang Zhao$^{1,2,5}$~\footnote{E-mail: zhaoq@ihep.ac.cn, Corresponding author} }

\affiliation{ 1) Institute of High Energy Physics and Theoretical Physics Center for Science Facilities,
        Chinese Academy of Sciences, Beijing 100049, China}

\affiliation{ 2)  University of Chinese Academy of Sciences, Beijing 100049, China}

\affiliation{ 3) Universit\'e de Lyon, Institut de Physique des 2 Infinis de Lyon, UCBL--IN2P3-CNRS,
4, rue Enrico Fermi, Villeurbanne, France}

\affiliation{ 4) Guangdong Provincial Key Laboratory of Nuclear Science,
Institute of Quantum Matter, South China Normal University, Guangzhou 510006, China}

\affiliation{ 5) Synergetic Innovation Center for Quantum Effects and
Applications (SICQEA), Hunan Normal University, Changsha 410081, China}

\begin{abstract}
The hadronic weak decays of $\Lambda_c$ are studied in the  framework of a constituent quark model. With the combined analysis of the Cabbibo-favored processes, $\Lambda_c\to \Lambda\pi^+$, $\Sigma^0\pi^+$ and $\Sigma^+\pi^0$, we confirm that the non-factorizable transition mechanisms play a crucial role in the understanding of their compatible branching ratios. We emphasize that the SU(3) flavor symmetry breaking effects, which is generally at the order of $1\sim 2\%$, can be amplified by the destructive interferences among the pole terms in the diagrams with internal conversion. Some contributions are sensitive to the spatial distribution of the scalar-isoscalar light-quark sector in the $\Lambda_c$, and its overlap with the light quarks in the final state hyperon. Namely, a compact diquark configuration is disfavored.

\end{abstract}
\date{\today}
%
\maketitle
\section{Introduction}
The hadronic weak decays of charmed baryons have served as a probe for QCD factorization. However, for a long time, due to the lack of precision measurements in experiments, crucial questions on the decay mechanisms have not been fully understood. In particular, it is not easy to calculate the  contributions from non-factorizable hadronic effects and evaluate the role played by the color suppressed processes.
Early theoretical studies of these processes based on different models can be found in the literature, for instance, algebraic techniques~\cite{Hussain:1983pk,Kaur:1991sw,Cheng:1991sn,Cheng:1993gf} which parameterized out typical amplitudes on the basis of symmetry considerations, and quark models~\cite{Korner:1978tc,Korner:1992wi,Uppal:1994pt} which calculate certain processes using explicit constituent wave functions. Interestingly, these prescriptions did not explicitly consider contributions from the color suppressed transitions, which were generally believed to be small.
In recent years other methods were applied to the study of the  hadronic weak decays of charmed baryons, such as the topological diagram approach~\cite{Zhao:2018mov}, QCD sum rules~\cite{Kisslinger:2018ciy} and spin-angular momentum structure analysis~\cite{Liang:2018rkl}. In addition, the weak decays of heavy baryons have been analyzed in the framework of SU(3) flavor symmetry ~\cite{Savage:1989qr,Sharma:1996sc,Lu:2016ogy,Geng:2019awr,Geng:2018rse,Geng:2018bow,Geng:2018plk, Geng:2017mxn}. Within this approach, one can relate all the relevant decay channels together and provide an overall systematic description of these processes. Predictions can then be made for those channels which have not yet been measured.
Initiated by the recent experimental progress on the $\Lambda_c$ decay measurements, the current-algebra approach is also used to revisit the $\Lambda_c$ decay in the MIT bag model~\cite{Cheng:2018hwl}. In this approach, the implementation of flavor symmetry is based on the  assumption of factorization, while the effects of non-factorizable processes are absorbed into some universal parameters. For the factorizable processes it is then assumed that the perturbative QCD (pQCD) should be the dominant dynamics.

Qualitatively, given that the mass of the charm quark is about 1.5\,GeV, it is not obvious that the decay of a charm quark into three light quarks should be dominated by the pQCD contributions, although the weak decay is generally a short-distance process. The  quarks emitted by the weak decay carry rather low momenta, thus, their hadronization should include significant  effects from final-state interaction. Namely, the color-suppressed transitions and pole terms both cannot be neglected if they are allowed by the quantum numbers. With the availability of high-precision measurements~\cite{Ablikim:2015flg,Ablikim:2017ors}, these controversial questions can be possibly addressed in an explicit quark model calculation. This motivates us to re-investigate the  hadronic weak decays of the charmed baryon $\Lambda_c$. Broader issues about the $\Lambda_c$ decays can be found in the recent literature. See, e.g., Refs.~\cite{Cheng:2015cca,Gronau:2018vei} and references therein.

As the first step for a systematic quark-model description, we study the two-body hadronic decays of $\Lambda_c$ into $\Lambda\pi$ and $\Sigma\pi$ which are the Cabbibo-favored processes. Our calculation includes both the factorizable process of  direct pion emission  and the processes that cannot be factorized.
The latter ones include the color-suppressed transitions and pole contributions due to the flavor internal conversion. By explicitly calculating these processes, we  demonstrate that their contributions cannot be neglected and their impact can provide useful insights into the effective constituent quark degrees of freedom in the quark model.

This paper is organized as follows. In Sec.~\ref{framework} the non-relativistic quark model framework is presented. The numerical results and discussions are given in Sec.~\ref{calculation}, and a brief summary is given in Sec.~\ref{summary}. In the Appendix, details are supplied for the quark wave functions and transition amplitudes.

\section {Framework}\label{framework}
\begin{figure}[ht]
\begin{center}
\includegraphics[scale=0.7]{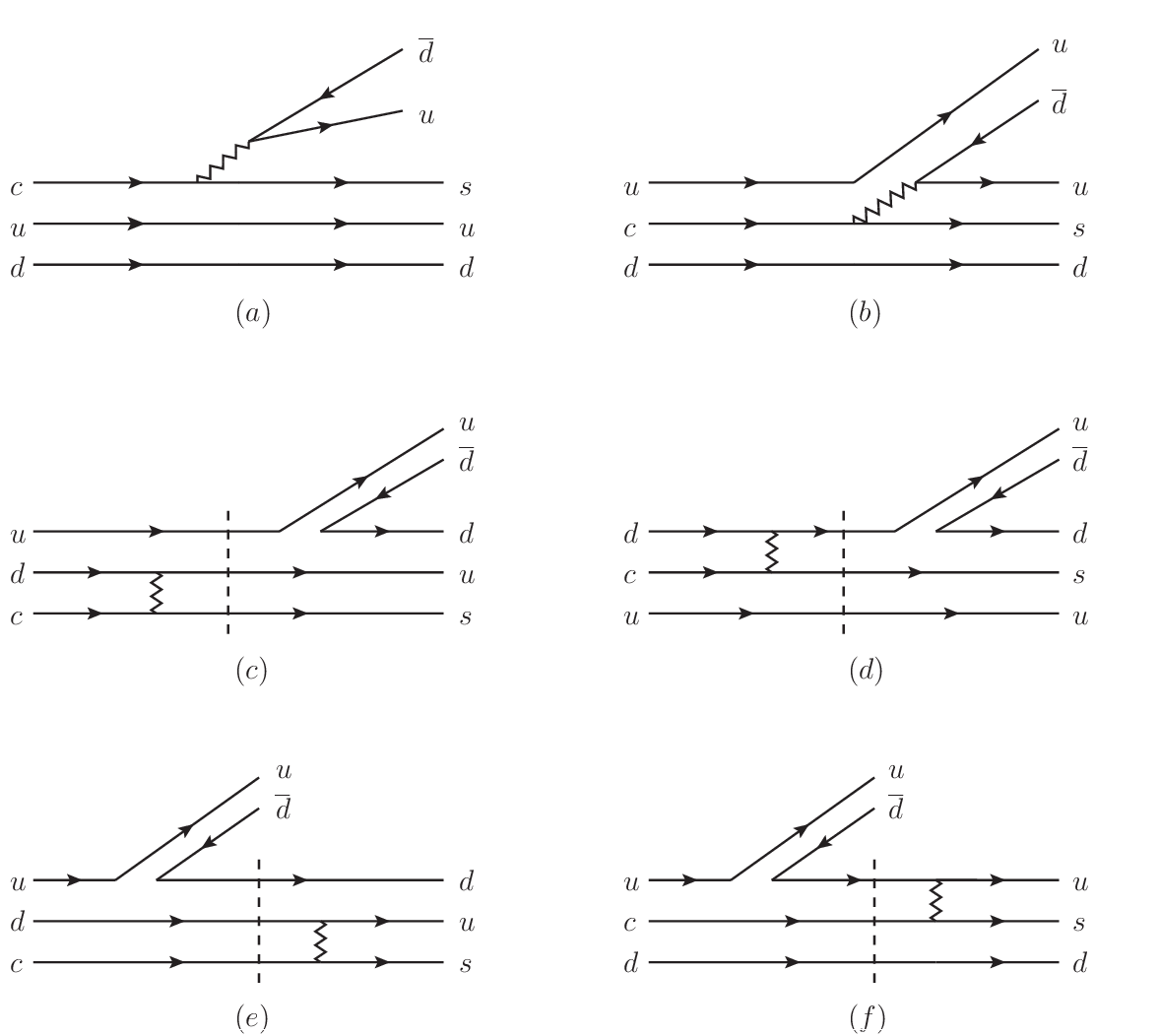}
\caption{Illustrations for the two-body hadronic weak decays of $\Lambda_c$ into $\Lambda\pi$ and $\Sigma \pi$ at the quark level.}
\label{fig:hadrondecay}
\end{center}
\end{figure}

In this paper we focus on the hadronic decays of $\Lambda_c\to \Lambda \pi^+$, $\Sigma^0 \pi^+ $, and $\Sigma^+ \pi^0$, which are all Cabbibo-favored processes. At leading order, there are two typical processes contributing to the weak pionic decays. One is the direct weak emission of a pion, and the other is the quark internal conversion inside the baryons. For the second type of processes, the pion is emitted by strong interaction vertices. The transitions involve the elementary weak transformations of $c \rightarrow s$ and $d\rightarrow u$ or $c \rightarrow s \bar d u$. These transition processes are illustrated in Fig.~\ref{fig:hadrondecay}, where (a) is the direct pion emission (DPE) process, (b) is the color suppressed (CS) pion emission, and (c)-(f) show the quark internal conversion processes. For Figs.~\ref{fig:hadrondecay} (c)-(f) the main contributions to these internal conversion processes should be via the intermediate pole terms. For these processes, the quantum numbers of intermediate baryon could be $\half^+$ for the parity-conserving (PC) process or $\half^-$ for the parity-violating (PV) one.

Some qualitative features can be learned from these transition processes. Figure~\ref{fig:hadrondecay} (a) is a typical factorizable process and calculable in pQCD. In contrast, all the other diagrams are non-factorizable and dominated by non-perturbative mechanisms.
If Fig.~\ref{fig:hadrondecay} (a) were the dominant decay mechanism for the $\Lambda_c$, the branching ratio for $\Lambda_c\to \Lambda\pi^+$ should be much larger than those for $\Lambda_c\to \Sigma\pi$, as the $ud$ pair is spectator.
However, the experimental branching ratios for $\Lambda_c\to \Lambda\pi^+$ and $\Sigma\pi$ are very similar, with branching ratios $(1.30\pm 0.07)\%$ for $\Lambda_c\to \Lambda\,\pi^+$, $(1.20\pm 0.07)\%$ for $\Lambda_c\to \Sigma^0\,\pi^+$ and $(1.25\pm 0.10)\%$ for $\Lambda_c\to \Sigma^+\,\pi^0$~\cite{Tanabashi:2018oca}. This is a strong evidence for the non-negligible contributions from those non-factorizable processes in Figs.~\ref{fig:hadrondecay} (b)-(f). By explicit calculations of these contributions in the quark model, it is interesting to compare the relative strengths among these amplitudes and learn about the roles played by the color suppressed (Fig.~\ref{fig:hadrondecay} (b)) and pole terms (Figs.~\ref{fig:hadrondecay} (c)-(f)).

\subsection{Convention}

Before proceeding to the detailed calculations, we  define the convention for the quark and antiquark fields:
\begin{equation}\begin{aligned}
q(x)&=\int \frac{d \bm p}{(2\pi)^{3/2}}\genfrac{(}{)}{}{}{m}{p^0}^{1/2} \sum_s \left[u_s(\bm p)b_s(\bm p)e^{i p\cdot x}+v_s(\bm p)d^\dagger_s(\bm p)e^{-i p\cdot x}\right], \\
\bar q(x)&=\int\frac{d\bm p}{(2\pi)^{3/2}}\genfrac{(}{)}{}{}{m}{p^0}^{1/2} \sum_s \left[\bar u_s(\bm p)b^\dagger_s(\bm p)e^{-i p\cdot x}+\bar v_s(\bm p)d_s(\bm p)e^{i p\cdot x}\right].
\end{aligned}
\end{equation}
The commutation and anticommutation relations of the creation and annihilation operators are given by:
\begin{align}
&\{b_s(\bm p),b_{s'}^\dagger(\bm p') \}=\{d_s(\bm p),d_{s'}^\dagger(\bm p') \}=\delta_{ss'}\delta^3(\bm p-\bm p').
\end{align}
The normalization of spinor is $u^\dagger_s(\bm p)u_{s'}(\bm p)=v^\dagger_s(\bm p)v_{s'}(\bm p)=(p^0/m)\delta_{ss'}$.  It should be noted that the spinor normalization must match the convention of the quark (anti-quark) field in order to keep the proper normalization of the quark (anti-quark) field.

In this work the mesons and baryons  are expressed with mock states~\cite{Hayne:1981zy}, respectively,
\begin{equation}\label{mock-meson}
\begin{alignedat}{2}
&
| M(\bm P_c)_{J,J_z}\rangle =&&\sum_{S_z,L_z;c_i}\langle L,L_z;S,S_z |J,J_z \rangle\int d \bm p_1 d \bm p_2 \delta^3(\bm p_1 +\bm p_2-\bm P_c)\Psi_{N,L,L_z}(\bm p_1 ,\bm p_2)\chi_{s_1,s_2}^{S,S_z} \\
&&&\times\frac{\delta_{c_1 c_2}}{\sqrt3} \phi_{i_1,i_2}b^\dagger_{c_1,i_1,s_1,\bm p_1}d^\dagger_{c_2,i_2,s_2,\bm p_2}|0\rangle, \\
&
| B(\bm P_c)_{J,J_z}\rangle =&&\sum_{S_z,L_z;c_i}\langle L,L_z;S,S_z |J,J_z \rangle\int d \bm p_1 d \bm p_2 d\bm p_3 \delta^3(\bm p_1 +\bm p_2+\bm p_3-\bm P_c) \Psi_{N,L,L_z}(\bm p_1 ,\bm p_2,\bm p_3) \chi_{s_1,s_2,s_3}^{S,S_z} \\
&&&\times\frac{\epsilon_{c_1 c_2 c_3}}{\sqrt6} \phi_{i_1,i_2,i_3} b^\dagger_{c_1,i_1,s_1,\bm p_1}b^\dagger_{c_2,i_2,s_2,\bm p_2}b^\dagger_{c_3,i_3,s_3,\bm p_3}|0\rangle,
\end{alignedat}
\end{equation}
where $c_j,\,s_j,\,i_j$ are color, spin, and flavor indexes, respectively; $\psi_{N,L,L_z}$ is the spatial wave function which is taken as an harmonic oscillator wavefunction; $\chi^{S,S_z}$ is the spin wave function; $\phi$ is the flavor wave function, and $\delta_{c_1c_2}/\sqrt 3$ and $\epsilon_{c_1c_2c_3}/\sqrt6$ are the color wave functions for the meson and baryon, respectively. The detailed expressions of these wave functions are given in Appendix~\ref{App:wavefuncion}. The normalization condition for the mock states are:
\begin{equation}\begin{aligned}
\langle  M(\bm P'_c)_{J,J_z} | M(\bm P_c)_{J,J_z}\rangle &=\delta^3(\bm P'_c-\bm P_c),\\
\langle  B(\bm P'_c)_{J,J_z} | B(\bm P_c)_{J,J_z}\rangle &=\delta^3(\bm P'_c-\bm P_c).\label{baryon-norm}
\end{aligned}
\end{equation}
In the above equations~\eqref{mock-meson}-\eqref{baryon-norm},  $\bm p_i$ denotes the single quark (antiquark) three-vector momentum, and $\bm P_c$ ($\bm P_c'$) denotes the hadron momentum.

Considering the two-body decay $A\to B+C$, the $S$ matrix in our framework is given by:
\begin{align}
S=I-2\pi i \delta^4(P_A-P_B-P_C)M,
\end{align}
with
\begin{align}
\delta^3(\bm P_A -\bm P_B-\bm P_C)M\equiv \langle BC | H_I|A \rangle.
\end{align}
Under this convention and by integrating over the phase space, the decay width is finally written as:
\begin{align}
\Gamma(A\to B+C)=8\pi^2\frac{|\bm k|E_B E_C}{M_A}\frac{1}{2 J_A+1} \sum_\text{spin}|M|^2,
\end{align}
where $\bm k$ is the three-momentum of the final state meson (e.g., the pion) in the initial state rest frame, $E_B$ and $E_C$ are the energies of the final-state particles $B$ and $C$, respectively, and $J_A$ is the spin of the initial state.
\subsection{Non-relativistic form of the effective Hamiltonian}
In this work we adopt a non-relativistic formalism. The weak decay probes the short-range dynamics inside hadrons, where a simple quark model is questionable. But we believe that most features of the short-range dynamics are parameterized and absorbed into the quark wavefunctions. Also, the hadronization involves long-distance dynamics, and it is consistently accounted for by the overlap of the initial- and final-state wavefunctions.
\subsubsection{Operators of the weak interaction}
The effective weak Hamiltonian (i.e., the form of four-fermion interactions) is generally written as \cite{Richard:2016hac,LeYaouanc:1988fx,LeYaouanc:1978ef}:
\begin{equation}
H_W=\frac{G_F}{\sqrt 2}\int d \bm x \frac12 \{ J^{-,\mu}(\bm x),J^{+}_{\mu}(\bm x) \},
\end{equation}
where
\begin{equation}
\begin{aligned}
J^{+,\mu}(\bm x)&=
\begin{pmatrix}\bar u&\bar c \end{pmatrix}
\gamma^\mu(1-\gamma_5)
\begin{pmatrix}\cos \theta_C & \sin \theta_C \\ -\sin \theta_C &\cos \theta_C \end{pmatrix} \begin{pmatrix} d\\s \end{pmatrix}, \\
J^{-,\mu}(\bm x)&=
\begin{pmatrix}\bar d &\bar c \end{pmatrix} \begin{pmatrix}\cos \theta_C & -\sin \theta_C \\ \sin \theta_C &\cos \theta_C \end{pmatrix} \gamma^\mu(1-\gamma_5)
\begin{pmatrix} u\\c \end{pmatrix}.
\end{aligned}
\end{equation}
According to its parity behavior under parity,  $H_W$ can be separated into a parity-conserving and a parity-violating part,
\begin{equation}
H_W=H^{PC}_W+H^{PV}_W, \notag
\end{equation}
where
\begin{equation}
\begin{aligned}
H_W^{PC}&=\frac{G_F}{\sqrt2}\int d\bm x\left[ j_\mu^-(\bm x)j^{+,\mu}(\bm x)+j_{5,\mu}^-(\bm x)j_5^{+,\mu}(\bm x) \right], \\
H_W^{PV}&=\frac{G_F}{\sqrt2}\int d\bm x\left[ j_\mu^-(\bm x)j_5^{+,\mu}(\bm x)+j_{5,\mu}^-(\bm x)j^{+,\mu}(\bm x) \right].
\label{eq:weakH}
\end{aligned}
\end{equation}
This Hamiltonian contains the tree-level operators and can be explicitly reduced into  non-relativistic forms for the $2\to 2$ internal conversion and $1\to 3$ emission processes, respectively. For the Cabbibo-favored $2\to 2$ quark transition process, the relevant term is
\begin{equation}
H_{W,2\to 2}=\frac{G_F}{\sqrt2}V_{ud}V_{cs} \frac{1}{(2\pi)^3}\delta^3(\bm p'_i+\bm p'_j-\bm p_i-\bm p_j) \bar u(\bm p_i')\gamma_\mu(1-\gamma_5)u(\bm p_i)\bar u(\bm p_j')\gamma^\mu(1-\gamma_5)u(\bm p_j).
\end{equation}
The creation and annihilation operators are omitted here and in the follow-up formulae. The non-relativistic expansion  gives:
\begin{equation}
\begin{aligned}
H_{W,2\to 2}^{PC}=&\frac{G_F}{\sqrt2}V_{ud}V_{cs}\frac{1}{(2\pi)^3} \sum_{i\neq j}\hat\alpha_i^{(-)} \hat\beta^{(+)}_j \delta^3(\bm p'_i+\bm p'_j-\bm p_i-\bm p_j)\left( 1- \langle s_{z,i}'|\bm \sigma_i|s_{z,i}\rangle \langle s_{z,j}'| \bm \sigma_j |s_{z,j}\rangle \right),
\\
%
H_{W,2\to 2}^{PV}=&\frac{G_F}{\sqrt 2} V_{ud}V_{cs}\frac{1}{(2\pi)^3}\sum_{i\neq j}\hat\alpha_i^{(-)}\hat\beta^{(+)}_j\delta^3(\bm p'_i+\bm p'_j-\bm p_i-\bm p_j) \\
&{}\times \left\{ -(\langle s_{z,i}'|\bm \sigma_i |s_{z,i}\rangle - \langle s_{z,j}'|\bm\sigma_j |s_{z,j}\rangle) \left [\left(\frac{\bm p_i}{2m_i}-\frac{\bm p_j}{2m_j}\right )+\left (\frac{\bm p'_i}{2m_i'}-\frac{\bm p'_j}{2m_j'}\right )\right ] \right.  \\
&{}+\left . i(\langle s_{z,i}'|\bm \sigma_i |s_{z,i}\rangle \times \langle s_{z,j}'|\bm\sigma_j |s_{z,j}\rangle) \left [\left (\frac{\bm p_i}{2m_i}-\frac{\bm p_j}{2m_j}\right )-\left (\frac{\bm p'_i}{2m_i'}-\frac{\bm p'_j}{2m_j'}\right)\right ] \right\},
\end{aligned}
\end{equation}
where $s_i$ and $m_i$ the spin and mass of the $i$-th quark, respectively; the subscripts $i$ and $j$ $(i,j=1,2,3 ~\text{and}~i\neq j)$ indicate the  quarks experiencing the weak interaction; $\hat\alpha_i$ and $\hat\beta_j$ are the flavor-changing operators, namely,
$\hat\alpha_i^{(-)}c_j=\delta_{ij}s_i,~\hat\beta_j^{(+)}d_i=\delta_{ij}u_i$; $V_{ud}$ and $V_{cs}$ are the Cabbibo-Kobayashi-Maskawa (CKM) matrix elements.
\begin{figure}[ht]
\begin{center}
\includegraphics[scale=0.25]{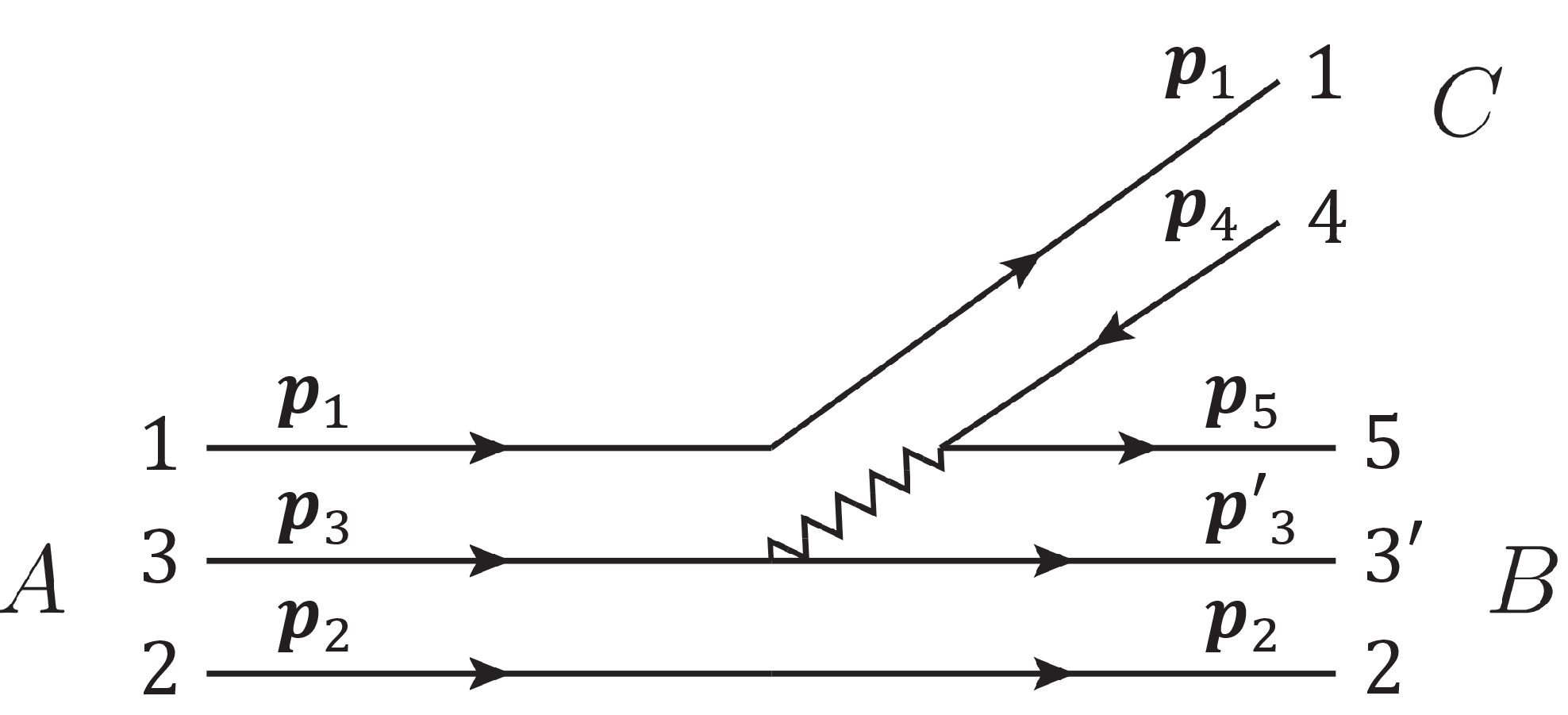}
\caption{The quark level diagram for the color suppressed transition process with quark labels.}
\label{fig:color}
\end{center}
\end{figure}

The $1\to 3$ transition operator can contribute to the direct pion emission and color suppressed processes. Figure~\ref{fig:color} illustrates the decay process of the CS at the quark level. In contrast, as for the DPE process, the light quarks $4$ and $5$ form the pion and the other final quarks form the baryon of the final states. Apart from the color factor, the different arrangements of quarks in the CS and DPE processes make a difference between these two processes. The calculation details will be given in next subsection. Here, we concentrate on the transition operator for Fig.~\ref{fig:color} which can be obtained with the explicit quark labels:
\begin{equation}
\begin{aligned}
H_{W,1\to 3}&=\frac{G_F}{\sqrt2}V_{ud}V_{cs}\frac{\beta}{(2\pi)^3}\delta^3(\bm p_3 -\bm p_3'-\bm p_5-\bm p_4 )
\bar u(\bm p_3',m_3')\gamma_\mu (1-\gamma_5)u(\bm p_3 ,m_3) \bar u(\bm p_5,m_5)\gamma^\mu (1-\gamma_5)v(\bm p_4 ,m_4)  \\
&=H_{W,1\to 3}^{PC}+H_{W,1\to3}^{PV},
\end{aligned}
\end{equation}
where $\beta$ is a symmetry factor. It takes a value of 3 in the DPE process and 2 in the CS process. The parity-conserving and the parity-violating parts are respectively written as
\begin{equation}
\begin{aligned}
H_{W,1\to 3}^{PC}
={}&\frac{G_F}{\sqrt2} V_{ud}V_{cs} \frac{\beta}{(2\pi)^3}\delta^3(\bm p_3 -\bm p_3'-\bm p_4-\bm p_5 ) \left\{ \la s_3'|I|s_3\ra \la s_5 \bar s_4|\bm \sigma|0\ra \left( \frac{\bm p_5}{2m_5}+ \frac{\bm p_4}{2m_4}\right) \right.\\
&{}-\left[ \left( \frac{\bm p_3'}{2m_3'}+\frac{\bm p_3}{2m_3} \right)\la s_3'|I|s_3\ra -i \la s_3'|\bm \sigma |s_3\ra \times \left( \frac{\bm p_3}{2m_3}-\frac{\bm p_3'}{2m_3'} \right) \right] \la s_5 \bar s_4|\bm \sigma|0\ra  \\
&{} -\la s_3'|\bm \sigma |s_3\ra \left[\left( \frac{\bm p_5}{2m_5}+\frac{\bm p_4}{2m_4} \right) \la s_5 \bar s_4 |I|0\ra
-i  \la s_5 \bar s_4 |\bm \sigma|0\ra\times \left( \frac{\bm p_4}{2m_4}-\frac{\bm p_5}{2m_5} \right)\right] \\
&{}+\la s_3'|\bm \sigma |s_3\ra  \left( \frac{\bm p_3'}{2m_3'}+\frac{\bm p_3}{2m_3} \right) \la s_5 \bar s_4 |I|0\ra \biggr\} \hat\alpha^{(-)}_3 \hat I'_\pi,\\
H_{W,1\to 3}^{PV}
={}&\frac{G_F}{\sqrt2} V_{ud}V_{cs} \frac{\beta}{(2\pi)^3}\delta^3(\bm p_3-\bm p_3'-\bm p_4-\bm p_5 )
\left( -\la s_3'|I|s_3\ra \la s_5 \bar s_4|I|0\ra + \la s_3'|\bm \sigma|s_3\ra \la s_5 \bar s_4|\bm \sigma|0\ra \right) \hat\alpha^{(-)}_3 \hat I'_\pi ,
\end{aligned}
\end{equation}
where $\bar s_4$ stands for the spin of particle $4$ which is an anti-quark.  In order to evaluate the spin matrix element including an anti-quark the particle-hole conjugation~\cite{Racah:1942gsc} should be employed. With the particle-hole conjugation relation $|j,-m\rangle\to(-1)^{j+m}|j,m\rangle$, the anti-quark spin transforms as follows: $\langle \bar \uparrow|\to|\downarrow\rangle$ and $\langle \bar \downarrow|\to -|\uparrow\rangle$. $I$ is the dimension-two unit matrix; $\hat\alpha^{(-)}$ is the flavor operator which transforms $c$ quark to $s$ and $\hat I'_\pi$ is the isospin operator for the pion production process. It has the form of
\begin{align}
\hat I'_\pi=\begin{dcases}
b^\dag_u b_u                    &\text{for}~ \pi^+,\\
-\frac{1}{\sqrt 2}b^\dag_u b_d   &\text{for}~ \pi^0,\\
\end{dcases}
\end{align}
for Cabbibo-favored processes and will act on the $i$-th quark of the initial baryon after considering the pion flavor wave function. As for the direct pion emission process, it is also a $1\to3$ weak interaction process. The operator for this process has the same form as for the color suppressed process except for the symmetry factor and delta functions. Without causing ambiguities the operators for both $2\to 2$ and $1\to3$ processes are labeled as $H_W$. Their differences are taken into account in the detailed calculations.
\subsubsection{Quark-meson couplings in the chiral quark model}
\begin{figure}[ht]
\begin{center}
\includegraphics[scale=0.7]{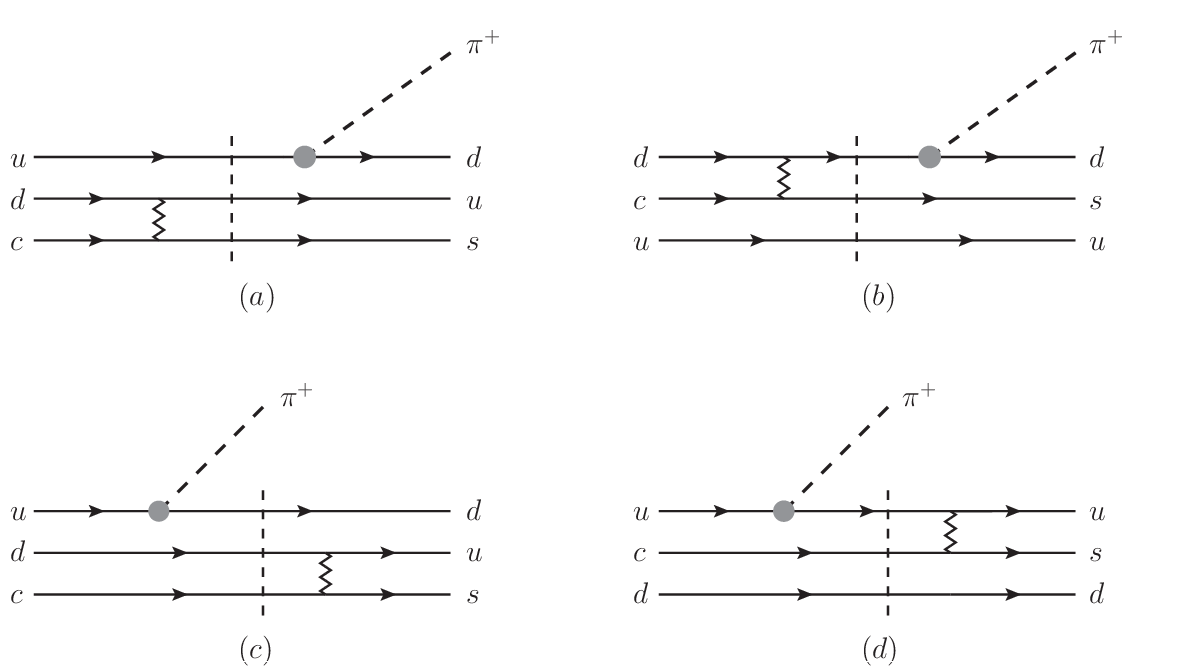}
\caption{The internal flavor conversion processes. The solid circle stands for the quark-pion vertex.}
\label{fig:chiralpic}
\end{center}
\end{figure}

For the production of a pion in the internal flavor conversion processes, the intermediate baryon pole terms become dominant. This allows an implementation of the chiral quark model~\cite{Manohar:1983md} for the pion production via the strong interaction vertices. The chiral quark model has been often applied to the production of light pseudoscalar mesons in various processes~\cite{Li:1997gd,Zhao:2002id,Zhong:2007gp}. In the chiral quark model the pion is treated as a fundamental particle. This treatment will simplify the calculations of processes in Figs.~\ref{fig:hadrondecay} (c)-(f) by their equivalence of Fig.~\ref{fig:chiralpic}.

The tree-level quark-meson pseudovector coupling can be deduced from the chiral quark model~\cite{Manohar:1983md} and the Hamiltonian can be written as:
\begin{align}
H_m=\sum_j \int d \bm x \frac{1}{f_m}\bar q_j(\bm x) \gamma_\mu^j \gamma_5^j q_j(\bm x) \partial^\mu \phi_m(\bm x)    \label{equ:Hpi},
\end{align}
where $f_m$ is the pseudoscalar meson decay constant; $q_j(\bm x)$ is the $j$-th quark field in the baryon and $\phi_m$ represents the meson field. In the non-relativistic limit the above equation can be expanded in the momentum space as:
\begin{align}
H_m=\frac{1}{\sqrt{(2\pi)^3 2\omega_m}}\sum_j \frac{1}{f_m}\left[\omega_m\left(\frac{\bm\sigma\cdot \bm p^j_f}{2m_f}+\frac{\bm\sigma\cdot \bm p^j_i}{2m_i}\right) -\bm \sigma \cdot \bm k\right] \hat I^j_m \delta^3(\bm p^j_f+\bm k-\bm p_i^j),
\end{align}
where $\omega_m$ and $\bm k$ are the energy and momentum of the pseudoscalar meson in the rest frame of the initial state, respectively; $\bm p^j_i$ and $\bm p^j_f$ are the initial and final momentum of the $j$-th quark, respectively; and $\hat I^j_m$ is the corresponding isospin operator for producing the pseudoscalar via its interaction with the $j$-th active quark within the baryon. For the production of the pion the isospin operator is written as:
\begin{align}
\hat I^j_\pi=\begin{dcases}
b^\dag_u b_d   &\text{for}~ \pi^-,\\
b^\dag_d b_u   &\text{for}~ \pi^+,\\
\frac{1}{\sqrt2}\left[b^\dag_u b_u- b^\dag_d b_d \right]    &\text{for}~ \pi^0,
\end{dcases}
\end{align}
where $b^\dag_{u,d}$ and $b_{u,d}$ are the creation and annihilation operators for the $u$ and $d$ quarks.

\subsection{Amplitudes}

In this section, we formulate the charmed-baryon decays with the operators and wave functions provided in the previous sections. The relevant transition processes have been given in Fig.~\ref{fig:hadrondecay}. For convenience we label the initial charmed baryon and final baryon as $B_c(\bm P_i;J_i,J^z_i)$ and $B_f(\bm P_f;J_f,J^z_f)$, respectively. The pion is labeled as $M_\pi(\bm k)$. Our calculation is performed in the rest frame of $\Lambda_c$, thus we have $\bm P_f=-\bm k$.

At the tree level the non-relativistic operators can be written as the following form
\begin{align}
H_I\equiv C \sum_n \hat O_n,
\end{align}
where $C$ is an overall factor and $\hat O_n$ is the direct product of flavor, spin and spatial operators:
\begin{equation}
\hat O_n=\hat O^\text{flavor}\hat O_n^\text{spin}\hat O_n^\text{spatial}.
\end{equation}
The transition matrix element can then be calculated in the quark model:
\begin{equation}\begin{aligned}
\langle B'(\bm P_f;J_f,J^z_f) |\hat O_n|&B(\bm P_i;J_i,J^z_i) \rangle \\
&{}=\sum_{S_f^z,L_f^z;S_i^z,L_i^z}[\langle \phi_f |\hat O^\text{flavor}|\phi_i \rangle \langle \chi_f^{S_f,S_f^z} |\hat O_n^\text{spin}|\chi_i^{S_i,S_i^z} \rangle \langle \Psi_f^{N_f,L_f,L_f^z}|\hat O_n^\text{spatial}|\Psi_i^{N_i,L_i,L_i^z} \rangle],
\end{aligned}
\end{equation}
where $J^z_{f/i}=S^z_{f/i}+L^z_{f/i}$ and $\sum_{S_f^z,L_f^z;S_i^z,L_i^z}[\cdots]$ is a shorthand notation for the Clebsch-Gordan sum; $\Psi$, $\chi$, $\phi$ denote the spatial, spin, and flavor wave functions, respectively, in the non-relativistic quark model~\cite{Copley:1979wj,Isgur:1978xj}. Also, we take the SU(6) spin-flavor wavefunctions in the calculation. It should be noted that in reality the SU(6) spin-flavor symmetry is broken due to the spin-dependent interactions. But as discussed in the literature~\cite{Isgur:1978xj,Capstick:2000qj} the low-lying baryons can still be reasonably described by the SU(6) wavefunctions as the leading approximation. In the processes of interest here the quark model uncertainties appear as an overall effect and can be absorbed into the quark model parameters. By adopting the SU(6) wavefunctions for the final state light baryons we can significantly simplify the calculations with the main conclusions intact.
\subsubsection{Amplitudes of the direct and color suppressed pion emission processes}
We now present some details on how to calculate these matrix elements in our framework. The DPE shown in Fig.~\ref{fig:hadrondecay} (a) can be expressed as:
\begin{align}
\label{eq:spatial}
M_{DPE}^{J_f,J_f^z;J_i,J_i^z}&=\langle B_f(\bm P_f;J_f,J_f^z)M(\bm k)|H_{W,1\to3}|B_c(\bm P_i;J_i,J_i^z) \rangle.
\end{align}
For the DPE process, the momentum conservation requires  $\bm P_f=\bm p_1+\bm p_2+\bm p_{3^\prime}$ and $\bm k=\bm p_5+\bm p_4$. This is guaranteed by the delta function in Eq.~(\ref{eq:IDPE}) with the spatial wave functions included.
The calculation of flavor and spin part can be found in~\cite{LeYaouanc:1988fx,Ackleh:1996yt}. The general form of the spatial wave function convolution that appears in the calculation for the DPE is written as
\begin{equation}
\begin{aligned}
\label{eq:IDPE}
I_{DPE}^{L_f,L_f^z;L_i,L_i^z}={}
&\langle \psi_\pi(\bm k) \Psi_{N_f, L_f, L^z_f}(\bm P_f) |\hat O_{W,1\to3}^\text{spatial}(\bm p_i)|\Psi_{N_i, L_i,L^z_i}(\bm P_i)\rangle\\
={}&\int d \bm p_1 d \bm p_2 d \bm p_3 d \bm p_{3^\prime} d \bm p_4 d \bm p_5
\Psi_{N_f,L_f,L_f^z}^*(\bm p_1,\bm p_2,\bm p_{3^\prime})\delta^3(\bm P_f-\bm p_1-\bm p_2-\bm p_{3^\prime})  \\
&{}\times\Psi_{0,0,0}^*(\bm p_4,\bm p_5)\delta^3(\bm k-\bm p_5-\bm p_4)  \hat O_{W,1\to3}^\text{spatial}(\bm p_i) \Psi_{N_i,L_i,L_i^z}(\bm p_1,\bm p_2,\bm p_3)\delta^3(\bm P_i-\bm p_1-\bm p_2-\bm p_3) \\
&{}\times \delta^3(\bm p_3-\bm p_4-\bm p_5-\bm p_{3^\prime}) ,
\end{aligned}
\end{equation}
where $\hat O_{W,1\to3}^\text{spatial}(\bm p_i)$ is the function of quark momentum $\bm p_i$, such as ${\bm p_5}/({2m_5})+{\bm p_4}/({2m_4})$ or just $1$ for $H_{W,1\to3}^{PV}$.

Since the DPE process is factorizable, its amplitude can also be written as:
\begin{align}
\label{eq:fac}
M_{DPE}^{J_f,J_f^z;J_i,J_i^z}=\frac{G_F}{\sqrt2}V_{ud}V_{cs} \la M(\bm k)_{\pi^+}\middle| \bar u\gamma^\mu(1-\gamma_5) d \middle| 0 \ra \la B_f(\bm P_f,J_f,J_f^z) \middle| \bar s\gamma_\mu(1-\gamma_5)c\middle| B_i(\bm P_i,J_i,J_i^z) \ra
\end{align}
where the pion creation is described by the axial current via
\begin{align}
\la M(\bm k)_{\pi^+}\middle| \bar u \gamma_5 \gamma^\mu d \middle| 0 \ra =i f_\pi p^\mu,
\end{align}
where $p^\mu$ is the four momentum of $\pi^+$ and $f_\pi$ is the pion decay constant. This form indicates that the DPE term is proportional to the pion momentum. In the hadronic weak decays of light octet baryons, the contribution from the DPE is much smaller than those from the pole terms~\cite{LeYaouanc:1988fx}. This can be understood by the relatively large momentum carried by the emitted pion and relatively large suppression from the off-shell pole propagators. Within our framework, by distinguishing the pole terms, we  describe the color-suppressed processes as contributions from the local current-current interactions that directly produce the pion after the weak transition. This allows us to compare the contributions between the DPE and CS processes.

The expression of the CS amplitude is similar to that of the DPE process:
\begin{equation}
M_{CS}^{J_f,J_f^z;J_i,J_i^z}=\langle B_f(\bm P_f;J_f,J_f^z)M(\bm k)|H_{W,1\to3}|B_c(\bm P_i;J_i,J_i^z) \rangle,
\end{equation}
Note that, for the CS process the momentum conservation requires $\bm P_f=\bm p_5+\bm p_2+\bm p_{3^\prime}$ and $\bm k=\bm p_1+\bm p_4$, which is different from the case of DPE. The spatial integral has the following expression:
\begin{equation}\begin{aligned}
I_{CS}^{L_f,L_f^z; L_i,L_i^z}={}&\langle \psi_\pi(\bm k) \Psi_{N_f, L_f, L^z_f}(\bm P_f)|\hat O_{W,1\to3}^\text{spatial}(\bm p_i)|\Psi_{N_i, L_i, L^z_i}(\bm P_i) \rangle\\
{}={}&\int d \bm p_1 d \bm p_2 d \bm p_3 d \bm p_{3^\prime} d \bm p_4 d \bm p_5
\Psi_{N_f, L_f, L_f^z}^*(\bm p_5,\bm p_2,\bm p_{3^\prime})\delta^3(\bm P_f-\bm p_5-\bm p_2-\bm p_{3^\prime}) \\
&{}\times\Psi_{0,0,0}^*(\bm p_1,\bm p_4)\times\delta^3(\bm k-\bm p_1-\bm p_4) \hat O_{W,1\to3}^\text{spatial}(\bm p_i) \Psi_{N_i, L_i, L_i^z}(\bm p_1,\bm p_2,\bm p_3)\delta^3(\bm P_i-\bm p_1-\bm p_2-\bm p_3) \\
&{}\times\delta^3(\bm p_3-\bm p_4-\bm p_5-\bm p_{3^\prime}).
\end{aligned}\end{equation}

It is interesting to analyze the differences between these two integral functions  $I_{DPE}^{L_f,L_f^z;L_i,L_i^z}$ and $I_{CS}^{L_f,L_f^z;L_i,L_i^z}$. For these two processes, apart from the $1/N_c$ suppression on the CS process, where $N_c$ is the number of colors, the difference between the spatial configurations in their wavefunction convolutions reflects the difference caused by the quark correlations. Note that the branching ratios for $\Lambda_c\to \Lambda\pi^+$, $\Sigma^0 \pi^+$ and $\Sigma^+\pi^0$ are at the same order of magnitude. It implies the importance of non-factorizable mechanisms which should become non-negligible in all these decay processes. Nevertheless, a coherent description of these processes can also provide hints on the nature of the light $ud$ diquark structure.

\subsubsection{Amplitudes of baryon internal conversion processes}

The baryon internal conversion processes shown in Fig.~\ref{fig:hadrondecay} (c)-(f) or Fig.~\ref{fig:chiralpic} are also called pole terms. They are two-step processes with the baryon weak transition either  preceding or following the strong pion emission. Because of the symmetry of the wave function, the processes shown by Fig.~\ref{fig:chiralpic} (a)-(b) (labeled as A-type pole terms) or (c)-(d) (labeled as B-type pole terms) can be included in one of the operators given in the previous section. Taking A-type process as an example, we can write the the amplitude for the baryon internal conversion processes as
\begin{equation}
M_\text{Pole,A}^{J_f,J_f^z;J_i,J_i^z}=M_{\text{Pole,A};PC}^{J_f,J_f^z;J_i,J_i^z}+M_{\text{Pole,A};PV}^{J_f,J_f^z;J_i,J_i^z},
\end{equation}
where
\begin{equation}
\begin{aligned}\label{equ:amp}
&\begin{multlined}
M_{\text{Pole,A};PC}^{J_f,J_f^z;J_i,J_i^z} \\
=\la B_f(\bm P_f;J_f,J_f^z) \middle| H_\pi \middle| B_m(\bm P_i;J_i,J_i^z) \ra \frac{i}{ \slashed p_{B_m}-m_{B_m} +i \frac{\Gamma_{B_m}}{2} } \la B_m(\bm P_i;J_i,J_i^z) \middle| H^{pc}_{W,2\to2} \middle| B_c(\bm P_i;J_i,J_i^z) \ra
\end{multlined}
\\
&\begin{multlined}
M_{\text{Pole,A};PV}^{J_f,J_f^z;J_i,J_i^z} \\
= \la B_f(\bm P_f;J_f,J_f^z) \middle| H_\pi  \middle| B'_m(\bm P_i;J_i,J_i^z) \ra  \frac{i}{ \slashed p_{B'_m}-m_{B'_m} +i \frac{\Gamma_{B'_m}}{2} } \la B'_m(\bm P_i;J_i,J_i^z) \middle|  H^{pv}_{W,2\to2} \middle| B_c(\bm P_i;J_i,J_i^z) \ra,
\end{multlined}
\end{aligned}
\end{equation}
in which $\left. \middle|B_m(\bm P_i;J_i,J_i^z)\ra$ and $\left. \middle|B'_m(\bm P_i;J_i,J_i^z)\ra$ denote the intermediate baryon states of $J^P=\half^+$ and $\half^-$, respectively, and $H_\pi$ means $\hat I^j_\pi$ is taken for $H_m$. In principle, all possible intermediate baryons, namely resonances and continuum states, should be included as the intermediate pole contributions for both parity conserved and parity violated processes~\cite{Cheng:2018hwl}. However, the main contributions come from the intermediate states  with low orbital momentum and  energy close to their on-shell mass.  For this reason, we only consider in this study the ground states and first orbital excitations.

For the intermediate baryon states, the non-relativistic form for their propagators is applied:
\begin{align}
\label{eq:propagator}
\frac{1}{\slashed p-m+ i\Gamma/2}\cong \frac{2m }{p^2-m^2+i \Gamma m}.
\end{align}
It should be cautioned that this treatment will bring uncertainties into the theoretical results since the intermediate states are generally off-shell. However, such uncertainties can be absorbed into the quark model parameters for which the range of the favored values by experimental data can be estimated.

Then, the parity conserved transition matrix element $\langle B(\bm p')| H_{W,2\to2}^{pc} |B_c(\bm p) \rangle$ can be directly expressed as, considering the simplified form of $ H_{W,2\to2}^{pc}$,
\begin{equation}\label{intern-conv-01}
\begin{split}
\langle B(\bm P')| H_{W,2\to2}^{pc} |B_c(\bm P)\rangle
{}={}&\frac{G_F}{\sqrt2}V_{ud}V_{cs} \frac{6}{(2\pi)^3}\int d\bm p_1 d\bm p_2 d\bm p_3\int d\bm p'_1 d\bm p'_2 d\bm p'_3
\delta^3(\bm p'_1+\bm p'_2-\bm p_1-\bm p_2)\delta^3(\bm p'_3-\bm p_3)  \\
&{}\times\Phi^*(\bm p'_1,\bm p'_2,\bm p'_3) \hat\alpha_1^{(-)} \hat\beta^{(+)}_2 \left( 1-\bm \sigma_1 \cdot \bm \sigma_2 \right)
\Phi(\bm p_1,\bm p_2,\bm p_3),
\end{split}
\end{equation}
where $\Phi(\bm p_1,\bm p_2,\bm p_3)$ and $\Phi(\bm p'_1,\bm p'_2,\bm p'_3)$ are the total wave function of the initial and final state baryon, respectively. Because of the symmetry of the total wave function, we can fix the subscript $i$ and $j$ to be $1$ and $2$ to compute the transition matrix element. The final amplitude will equal to the result multiplied by a symmetry factor $6$. Similarly, as we did before, we can obtain the transition matrix element $\langle B(\bm P')| H_{W,2\to2}^{pv} |B_c(\bm P) \rangle$ and $\langle B(\bm P')| H_\pi |B_c(\bm P) \rangle$.

\section{Numerical Results and discussions}
\label{calculation}
\subsection{Parameters and inputs}
Before presenting the numerical results, we clarify the parameters and inputs in our calculation as follows:

We adopt the same value $m_q=0.35\,$GeV for the masses of the $u,\, d$ and $s$ quarks. Taking the same mass for both nonstrange and strange quarks means that we take the SU(3) flavor symmetry as a leading approximation. Accordingly, we describe the light baryon with the same oscillator parameters $\alpha'_\lambda=\alpha'_\rho=0.4\,$GeV which is consistent with Refs.~\cite{Zhong:2007gp,Nagahiro:2016nsx}. This treatment is based on an empirical consideration of compromising the model uncertainties and simplifications. In the nonrelativistic quark model SU(3) flavor symmetry breaking effects explicitly appear in the eignvalues of the Hamiltonian via the mass term and mass-dependence in the kinetic energy and in the potential. Meanwhile, the harmonic oscillator strength for the correlations between the non-strange and strange quarks will also be different from that for the non-strange quarks. In Ref.~\cite{Isgur:1978xj} the harmonic oscillator strength difference between $s=0$ and $s=-1$ states due to the SU(3) symmetry breaking is expressed as $\omega_\rho-\omega_\lambda=\omega [1-\sqrt{(2x+1)/3}]$ with $x\simeq m_{u/d}/m_s=0.6$ is adopted. However, as shown by Ref.~\cite{Isgur:1978xj} and later calculations (see review of Ref.~\cite{Capstick:2000qj}), the harmonic oscillator strength difference is actually small. With $\omega_\rho=\omega_\lambda$ in the equal-mass treatment the same quality in the description of low-lying light baryons can be achieved. This indicates that the SU(3) flavor-symmetry breaking effects on the baryon masses are leading order contributions but are subleading ones on the wavefunctions. It leaves the leading SU(3) flavor-symmetry breaking effects to be manifested by the differences among baryon masses in the pole terms, and allows us to make the approximation of adopting the physical masses and widths in the propagators for the intermediate states.

We take the charm quark mass $m_c=1.5\,$ GeV and adopt for the wave function of the charmed baryon the parameters $\alpha_\rho=0.45\,$ GeV and $\alpha_\lambda=[{3m_c}/(2m_q+m_c)]{}^{1/4}\alpha_\rho$. The explicit expressions are given in Appendix~\ref{App:wavefuncion}. The pion wave function is also expressed as a Gaussian with a parameter $R=0.28$ GeV. Since the pion is extremely light and associated with the spontaneous chiral symmetry breaking, our treatment is empirical and some intrinsic uncertainties are unavoidable. However, we would like to stress that the effects arising from the pion wave function can be examined by varying the parameter $R$ within a reasonable range.

The intermediate states of the pole terms contribute differently in these three decay processes. To be more specific, we note that both $\Sigma^+$ and $\Sigma^{*+}$ will contribute to the A-type pole terms of all three decays. In contrast, $\Sigma_c^0$ and $\Sigma_c^{*0}$ will contribute to the B-types pole terms in $\Lambda_c\to\Lambda\pi^+$ and $\Lambda_c\to\Sigma^0\pi^+$. For the intermediate states in $\Lambda_c\to\Sigma^+\pi^0$  one notices that both $\Sigma_c^+$ and $\Sigma_c^{*+}$ can contribute. In our calculation the intermediate states of pole terms are as follows:
\begin{itemize}
\item $\Sigma^+ \ (1/2^+)$, $\Sigma^{*+}(1620) \ (1/2^-)$ and $\Sigma^{*+}(1750) \ (1/2^-)$ for the A-type pole terms in all three channels;
\item $\Sigma_c^0 \ (1/2^+)$ and $\Sigma_c^{*0} \ (1/2^-)$ for the B-type pole terms in $\Lambda_c\to\Lambda\pi^+$ and $\Lambda_c\to\Sigma^0\pi^+$;
\item $\Sigma_c^+ \ (1/2^+)$ and $\Sigma_c^{*+} \ (1/2^-)$ for the B-type pole terms in $\Lambda_c\to\Sigma^+\pi^0$.
\end{itemize}
Although the quantum numbers of  $\Sigma_c^*(2806)$ and $\Sigma_c^*(2792)$ as the first orbital excitation states with $J^P=1/2^-$ have not yet be measured in experiment, their masses are consistent with the quark model expectations~\cite{Copley:1979wj}. Their masses are adopted from the Particle Data Group~\cite{Tanabashi:2018oca} and  listed in Table~\ref{tab:mass}.

\begin{table}[ht]
  \centering
  \caption{ The baryon masses and widths taken from PDG~\cite{Tanabashi:2018oca} in the calculation. Only the central values of the masses and widths are listed. Note that the  $J^P=1/2^-$ states for both charmed and strange baryons have not been well determined. We assign $\Sigma_c^{*0}(2806)$ and $\Sigma_c^{*+}(2792)$ for the charmed states and $\Sigma^{*+}(1620)$ and $\Sigma^{*+}(1750)$ for the strange baryons with $J^P=1/2^-$. }
    \begin{ruledtabular}
    \renewcommand{\arraystretch}{1.2}
    \begin{tabular}{ccccccccccc}
 Particles&$\Lambda$     &$\Lambda_c$    &$\Sigma^0$      &$\Sigma^+$ &$\Sigma^{*+}(1620)$
          &$\Sigma^{*+}(1750)$ &$\Sigma_c^{0}$ &$\Sigma_c^{*0}$ &$\Sigma_c^{+}$ &$\Sigma_c^{*+}$\\
    \hline
 $I(J^P)$ &$0(\half^+)$&$0(\half^+)$&$1(\half^+)$&$1(\half^+)$&$1(\half^-)$
          &$1(\half^-)$&$1(\half^+)$&$1(\half^-)$&$1(\half^+)$&$1(\half^-)$\\
 Mass(GeV)&$1.116$           &$2.286$       &$1.193$       &$1.189$ &$1.62$
          &$1.75$&$2.453$       &$2.806$       &$2.452$       &$2.792$
 \\
 Width(GeV) &   -          &   -    &   -   &   -     &$0.050$
           &$0.050$ &$0.00183$    &$0.072$ &$0.0046$        &$0.062$\\
\end{tabular}\end{ruledtabular}
  \label{tab:mass}
\end{table}

For those transitions involving the intermediate pole terms the intermediate states are off-shell in the kinematic regions of consideration. We leave the off-shell effects to be described by the wave function convolutions which eventually play the role of an  interaction form factors. The internal conversion will then keep the energy and three-momentum conservation, respectively, as shown in Eq.~(\ref{intern-conv-01}). For instance, in Fig.~\ref{fig:chiralpic} (a) the amplitude for $\Sigma_c\to \Sigma$ is defined at the mass of $\Sigma_c$ which means that $E_\Sigma=m_{\Sigma_c}$ and ${\bm P}_\Sigma=0$ in the $\Sigma_c$ rest frame. The propagators also take off-shell values as required.

\subsection{Numerical results and analyses}

Comparing the decay channels of $\Lambda\pi$ and $\Sigma\pi$, one of the interesting features is that the $\Lambda\pi$ channel allows the direct pion emission while it is forbidden in the $\Sigma^0\pi^+$ channels. This can be directly recognized because the $ud$ quarks are spectators in the factorizable transitions where the $c$ quark decays into $s+ \pi^+$. Since the initial $ud$ diquark is in color $\bar{\bm 3}$ with $(I_{ud}, \ J_{ud})=(0, \ 0)$ the $\Lambda_c$ cannot decays into $\Sigma^0\pi^+$ via the DPE transition. For $\Lambda_c\to \Sigma^+\pi^0$ it is suppressed by the neutral current interaction. This makes the combined analyses of these three channels useful for disentangling the underlying mechanisms. Note that the experimental data for the branching ratios of these three channels are compatible. It suggests that the DPE process should not be the only dominant contribution and other transition mechanisms must be considered. This should be a direct evidence for the non-negligible role played by non-factorizable processes in the non-leptonic decays of $\Lambda_c$. Some detailed formulations are given in
Appendix~\ref{App:am}.

We also note that these three decay channels share a similar form for the pole terms and for the color suppressed term.
The reason is because the final state $\Lambda$ and $\Sigma$ belong to the same SU(3) flavor multiplet. Thus, their spatial wave functions are the same at the leading order of the SU(3) flavor symmetry. The amplitudes of the pole terms or color suppressed term will be distinguished by the flavor transition factor.
Note that the measured branching ratios of these two channels are almost the same. It indicates that they share the same mechanisms via the non-factorizable transitions.

\begin{table*}[ht]
\renewcommand{\arraystretch}{1.3}
\caption{The flavor matrix elements for the CS process.}
\begin{ruledtabular}\begin{tabular}{ccccc}
Processes &$\langle \phi^\lambda_\Sigma|\hat\alpha^{(-)}_3 \hat I'_{\pi,1}|\phi^\lambda_{\Lambda_c}\rangle$
&$\langle\phi^\lambda_\Sigma|\hat\alpha^{(-)}_3 \hat I'_{\pi,1}|\phi^\rho_{\Lambda_c}\rangle$
&$\langle\phi^\rho_\Sigma|\hat\alpha^{(-)}_3 \hat I'_{\pi,1}|\phi^\lambda_{\Lambda_c}\rangle$
&$\langle(\phi^\rho_\Sigma|\hat\alpha^{(-)}_3 \hat I'_{\pi,1}|\phi^\rho_{\Lambda_c}\rangle$  \\
\hline
$\Lambda_c\to\Sigma^0\pi^+$ & $0$ &$-\nicefrac{1}{3}$ &$0$ &$0$ \\
$\Lambda_c\to\Sigma^+\pi^0$ & $0$ &$-\nicefrac{1}{3}$ &$0$ &$0$
    \end{tabular}\end{ruledtabular}
     \label{tab:flavor-CS}
\end{table*}

Taking the color suppressed process as an example, the flavor transition elements are given Tab.~\ref{tab:flavor-CS}. The only nonvanishing element is $\langle \phi_{\Sigma^0}^\lambda|\hat\alpha{}^{(-)}_3 \hat I{}'_{\pi,1}| \phi_{\Lambda_{c}}^\rho\rangle$. Note that in the parity-violating process the contributing flavor operator is between the $\phi_{\Lambda_{c}}^\rho$ and $\phi_{\Sigma^0}^\lambda$ configurations. This means that the parity-violating amplitudes can actually probe the structure arising from the $ud$ diquark-type of  correlations in the initial $\Lambda_c$ wave function. For the parity-conserving process the nonvanishing transition matrix elements in the spin-flavor spaces are via $\rho\to \rho$ type of transitions (The $\lambda\to \lambda$ type is suppressed by the vanishing of the $\lambda$-type wave function in the initial $\Lambda_c$, if one adopts the quark model). These features will allow us to examine the $ud$ correlation effects by the combined analyses of these three channels.

In Tabs.~\ref{tab:spin-PC} and \ref{tab:spin-PV} the spin matrix elements for the parity-conserving and parity-violating operators are listed, respectively, for different spin configurations. Note that the nonvanishing transition matrix elements should combine the averaged values in both flavor and spin space.
\begin{table*}[ht]
  \renewcommand{\arraystretch}{1.3}
  \caption{The spin matrix elements for the parity-conserving transitions in the CS process. Note that the spin wave function of pion is omitted.}
 \begin{ruledtabular}\begin{tabular}{ccccc}
$\mathcal{O}^\text{spin}$
&$\langle\chi^\lambda_{\half,-\half}|\mathcal{O}^\text{spin}|\chi^\lambda_{\half,-\half}\rangle$
&$\langle\chi^\lambda_{\half,-\half}|\mathcal{O}^\text{spin}|\chi^\rho_{\half,-\half}\rangle$
&$\langle\chi^\rho_{\half,-\half}|\mathcal{O}^\text{spin}|\chi^\lambda_{\half,-\half}\rangle$
&$\langle\chi^\rho_{\half,-\half}|\mathcal{O}^\text{spin}|\chi^\rho_{\half,-\half}\rangle$ \\
\hline
$\langle s'_3|I|s_3 \rangle \langle s_5 \bar s_4|\sigma_z|0\rangle$
&$\dfrac{\sqrt2}{3}$ &$-\dfrac{1}{\sqrt{6}}$ &$-\dfrac{1}{\sqrt{6}}$ &$0$ \\
$\langle s'_3|\sigma_z|s_3 \rangle \langle s_5 \bar s_4|I|0\rangle$
&$-\dfrac{1}{3\sqrt2}$ &$0$ &$0$ &$\dfrac{1}{\sqrt{2}}$ \\
$(\langle s'_3|\bm \sigma|s_3 \rangle \times\langle s_5 \bar s_4|\bm \sigma |0\rangle)_z$  &$0$ &$\dfrac{2i}{\sqrt 6}$ &$-\dfrac{2i}{\sqrt 6}$ &$0$ \\
    \end{tabular}\end{ruledtabular}
\label{tab:spin-PC}
\end{table*}

\begin{table*}[ht]
  \begin{center}
  \renewcommand{\arraystretch}{1.3}
  \caption{The spin matrix elements for the parity-violating transitions in the CS process. Note that the spin wave function of pion is omitted.}
\begin{ruledtabular}    \begin{tabular}{ccccc}
$\mathcal{O}^\text{spin}$
&$\langle\chi^\lambda_{\half,-\half}|\mathcal{O}^\text{spin}|\chi^\lambda_{\half,-\half}\rangle$
&$\langle\chi^\lambda_{\half,-\half}|\mathcal{O}^\text{spin}|\chi^\rho_{\half,-\half}\rangle$
&$\langle\chi^\rho_{\half,-\half}|\mathcal{O}^\text{spin}|\chi^\lambda_{\half,-\half}\rangle$
&$\langle\chi^\rho_{\half,-\half}|\mathcal{O}^\text{spin}|\chi^\rho_{\half,-\half}\rangle$  \\
\hline
$\langle s'_3|I|s_3 \rangle \langle s_5 \bar s_4|I|0\rangle$ &$-\frac{1}{\sqrt2}$ &$0$ &$0$ &$- \dfrac{1}{\sqrt2}$ \\
$\langle s'_3|\sigma_x|s_3 \rangle \langle s_5 \bar s_4|\sigma_x|0\rangle$
&$ \dfrac{\sqrt2}{3}$ &$ \dfrac{1}{\sqrt{6}}$ &$ \dfrac{1}{\sqrt{6}}$ &$0$ \\
$\langle s'_3|\sigma_y|s_3 \rangle \langle s_5 \bar s_4|\sigma_y|0\rangle$
&$ \dfrac{\sqrt2}{3}$ &$ \dfrac{1}{\sqrt{6}}$ &$ \dfrac{1}{\sqrt{6}}$ &$0$ \\
$\langle s'_3|\sigma_z|s_3 \rangle \langle s_5 \bar s_4|\sigma_z|0\rangle$
&$ \dfrac{\sqrt2}{3}$ &$ \dfrac{1}{\sqrt{6}}$ &$\dfrac{1}{\sqrt{6}}$ &$0$ \\
    \end{tabular}\end{ruledtabular}
     \label{tab:spin-PV}
    \end{center}
\end{table*}

Another feature distinguishing the factorizable DPE process and non-factorizable processes is that the amplitudes have different dependence on the pion wave function. As mentioned before, we introduce the pion wave function using harmonic oscillator in our calculation. Although this is a very coarse approximation, it demonstrates the relative amplitude strengths between the factorizable and non-factorizable transitions change in terms of the pion structure. As shown in Appendix~\ref{App:am},  the amplitude of the DPE process for $\Lambda_c\to\Lambda \pi^+$ is proportional to $R^{3/2}$. In contrast, the dependence of the non-factorizable terms on the $R$ in the color suppressed process is very different and more complicated. It means that the interference between the factorizable DPE process and non-factorizable processes is indeed a nontrivial issue that should be investigated.

In Tab.~\ref{tab:allresult} we show the calculated amplitudes for the transition element with $J_f^z=J_i^z=-1/2$ for each type of processes as a comparison. It shows that the parity-conserving amplitudes of the pole terms are larger than the parity-violating ones. Moreover, it shows that the interference between the A-type and the B-type processes are destructive. With the vertex couplings determined in the quark model this sign difference can be attributed to the signs of the propagators in these two types of processes. Further interferences are provided by the CS process for all these three channels. In the $\Lambda\pi^+$ decay channel the CS amplitude is further suppressed in comparison with the DPE amplitude, which is smaller than $1/N_c=1/3$. However, if one combines the pole terms which are also non-factorizable and color-suppressed, the $1/N_c$ suppression factor seems still to hold. It shows that the interferences between the factorizable DPE and non-factorizable processes lead to the compatible branching ratios for these three decay channels.

\begin{table*}[ht]
  \begin{center}
  \renewcommand{\arraystretch}{1.3}
  \caption{The amplitudes with $J_f^z=J_i^z=-1/2$ for different processes and the unit is $10^{-9}$ $\text{GeV}^{-1/2}$.  Amplitudes $A1(PV)$ and $A2(PV)$ are given by the parity-violating intermediate states $\Sigma^{*+}(1620)$ ($[{\bf 70}, {\bf ^2 8}]$) and $\Sigma^{*+}(1750)$ ($[{\bf 70}, {\bf ^4 8}]$), respectively. }
\begin{ruledtabular}\begin{tabular}{cccccccccc}
    Processes &$A(PC)$ &$A1(PV)$ &$A2(PV)$ &$B(PC)$ &$B(PV)$  &$CS(PC)$  &$CS(PV)$  &$DPE(PC)$  &$DPE(PV)$ \\
    \hline
    $\Lambda_c\to\Lambda \pi^+$ &$-16.50$&$0.74-0.023i$&$-2.57+0.10i$ &$22.33+0.021i$   &$-10.72-0.33i$    &$3.50$   &$-4.17$ &$-42.47$ &$24.07$ \\
    $\Lambda_c\to\Sigma^0\pi^+$ &$19.67$ &$-3.21+0.10i$ &$-2.23+0.090i$ &$-40.73-0.040i$  &$19.16+0.60i$    &$-6.04$  &$7.53$  &$0$       &$0$     \\
    $\Lambda_c\to\Sigma^+\pi^0$ &$19.64$ &$-3.15+0.098i$&$-2.19+0.088$ &$-40.65-0.10i$  &$19.28+0.52i$    &$-6.04$  &$7.51$
&$0$       &$0$
     \end{tabular}\end{ruledtabular}
     \label{tab:allresult}
    \end{center}
\end{table*}

\begin{figure}[ht]
\begin{center}
\includegraphics[scale=0.5]{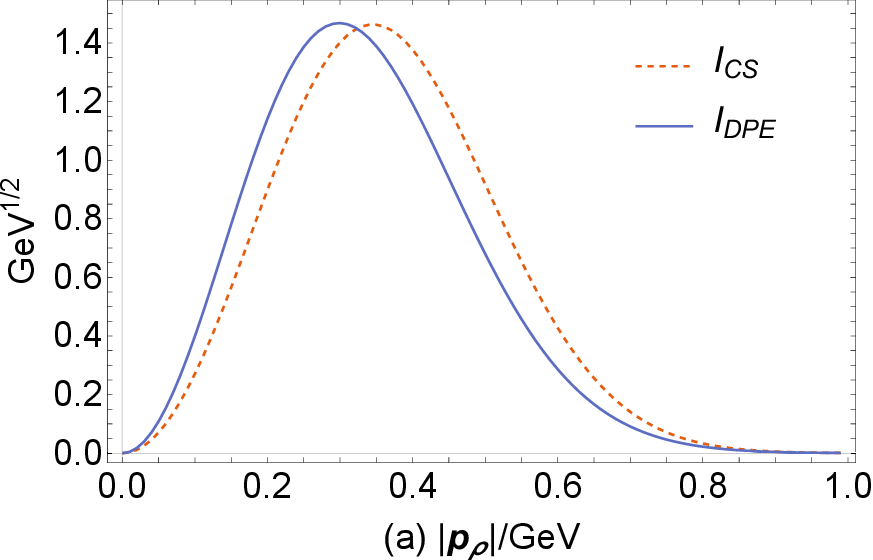}
\includegraphics[scale=0.5]{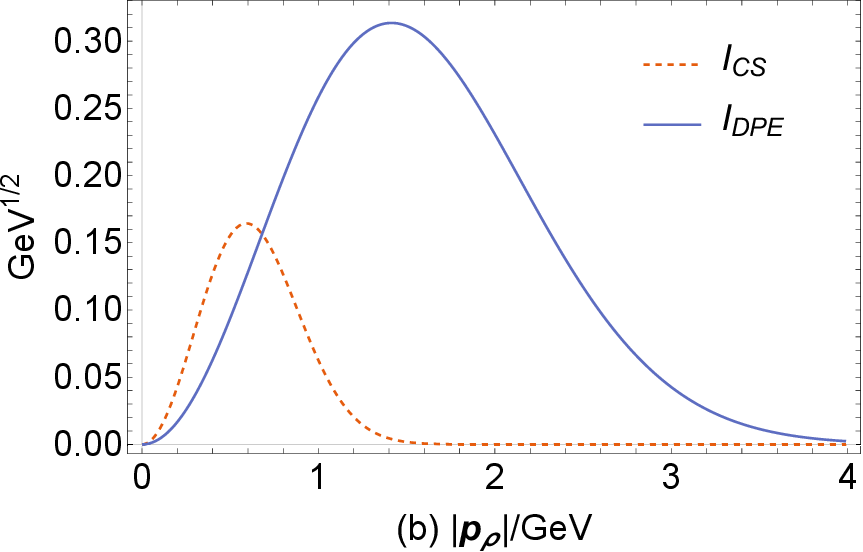}
\caption{(Colored) The spatial wave function convolutions of the DPE process (blue line) and CS process (brown line). The left panel shows the results with $\alpha_\rho=\alpha'_\rho=0.45$ GeV and the right one with $\alpha_\rho=\alpha'_\rho=2$ GeV.}
\label{fig:wfc}
\end{center}
\end{figure}

The phenomenological impact of the correlation among the light $ud$ quarks can be investigated here. It is obvious that the  convolution of the spatial wave functions depends on the structure of the hadrons that are involved. The question is whether there is a spatial correlation between the $u$ and $d$ quarks forming a compact structure, or simply a quantum-number correlation with their total spin and isospin 0. This can be examined by varying the parameter $\alpha_\rho$ of the wave function parameter which describes the relative distribution between $u$ and $d$. For small $\alpha_\rho$, one gets a loose Gaussian, and for large $\alpha_\rho$, one approaches a $\delta$-function.

For the transition processes of $\Lambda_c\to \Lambda\pi^+$, we can compare the spatial integrals for the DPE and CS processes and examine the $ud$ diquark correlations. The results are shown in Fig.~\ref{fig:wfc}. Note that the Fourier transformation of a Gaussian distribution function is still a Gaussian, we actually show the integrands in the momentum space with all the momenta except for $|\bm p_{\rho}|$ integrated out. Namely, we define functions $I_{CS}$ and $I_{DPE}$ as the results with all the momenta except $|\bm p_{\rho}|$ integrated out for $L_{CS}^{0,0;0,0}$ and $L_{DPE}^{0,0;0,0}$, and with the operator $O^{i,spatial}_{W,1\to3}\equiv 1$ in the Jacobi coordinate
\begin{align}
L_{CS}^{0,0;0,0}(\bm p_\rho) &=\int d\bm p_1 d\bm p_4 \int d\bm p_\lambda  d\bm p_\rho' d\bm p_\lambda' \delta^3(\bm k-\bm p_1-\bm p_4) \notag \\
&\times \delta^3(\frac{m_q}{M}\bm P_i+\frac{1}{2} \bm p_\lambda+\bm p_\rho-\bm p_1)\delta^3(\frac{m_q}{M}\bm P_i+\frac{1}{2} \bm p_\lambda-\bm p_\rho-\frac{m_q}{M'}\bm P_f-\frac{1}{2} \bm p'_\lambda + \bm p'_\rho)  \notag \\
&\times \delta^3(\frac{m_c}{M} \bm P_i-\bm p_\lambda -\frac{m_c}{M'} \bm P_f+\bm p'_\lambda -\frac{m}{M'}\bm P_f-\frac{1}{2} \bm p'_\lambda - \bm p'_\rho -\bm p_4)
\Psi_{0,0,0}(\bm p_\rho,\bm p_\lambda) \Psi^*_{0,0,0}(\bm p'_\rho,\bm p'_\lambda) \Psi^*_{0,0,0}(\bm p_1,\bm p_4),\\
L_{DPE}^{0,0;0,0}(\bm p_\rho) &=
\int d\bm p_4 d\bm p_5 \int  d\bm p_\lambda  d\bm p_\rho' d\bm p_\lambda'
\delta^3(\bm k-\bm p_4-\bm p_5)\notag \\
&\times \delta^3(\frac{m_q}{M}\bm P_i+\frac{1}{2} \bm p_\lambda+\bm p_\rho-\frac{m_q}{M'}\bm P_f-\frac{1}{2} \bm p'_\lambda - \bm p'_\rho) \delta^3(\frac{m_q}{M}\bm P_i+\frac{1}{2} \bm p_\lambda-\bm p_\rho-\frac{m_q}{M'}\bm P_f-\frac{1}{2} \bm p'_\lambda + \bm p'_\rho) \notag \\
&\times \delta^3(\frac{m_c}{M} \bm P_i-\bm p_\lambda -\frac{m_q}{M'} \bm P_f+\bm p'_\lambda -\bm k)
\Psi_{0,0,0}(\bm p_\rho,\bm p_\lambda) \Psi^*_{0,0,0}(\bm p'_\rho,\bm p'_\lambda) \Psi^*_{0,0,0}(\bm p_4,\bm p_5),
\end{align}
where
\begin{equation}
\left\{
\begin{aligned}
M&=2 m_q+m_c, \\[1pt]
\bm P_i&=\bm p_1+\bm p_2+\bm p_3,\\[1pt]
\bm p_\rho &=(\bm p_1-\bm p_2)/2,\\[1pt]
\bm p_\lambda &=(m_c\bm p_1+m_c\bm p_2-2m_q\bm p_3)/M ,
\end{aligned}
\right.
\end{equation}
and
\begin{equation}
\left\{
\begin{aligned}
M'=&3m_q, \\[1pt]
\bm P_f =& \bm p_5+\bm p_3'+\bm p_2,\\[1pt]
\bm p_\rho' =& (\bm p_5-\bm p_2)/2, \\[1pt]
\bm p_\lambda' =& (\bm p_5+\bm p_2-2\bm p_3')/3 \ .
\end{aligned}
\right.
\end{equation}

Figures~\ref{fig:wfc} (a) and (b) correspond to two  different values of $\alpha_\rho=0.45$ and 2\,GeV, respectively. As a further simplification we also take $\alpha'_\rho=\alpha_\rho$,  namely, the $ud$ pair with $(I_{ud}, \ J_{ud})=(0, \ 0)$ in the light baryon has the same spatial distribution as in the $\Lambda_c$. It shows that with the increase of $\alpha'_\rho=\alpha_\rho$, namely, if the $ud$ diquark becomes more compact, the CS contribution will be significantly suppressed compared to the DPE. In another word, the present experimental measurement favors that the correlation between the $ud$ diquark to be as extended as a conventional hadron size instead of a compact structure. Otherwise, the branching ratio for the $\Lambda\pi^+$ channel would be much larger than that for $\Sigma\pi$.

\begin{table*}[ht]
  \begin{center}
  \renewcommand{\arraystretch}{1.3}
  \caption{The calculated branching ratios (in \%) of the $\Lambda_c$ decays in this work are compared with experimental data~\cite{Tanabashi:2018oca,Ablikim:2015flg} and other model calculations~\cite{Geng:2017esc,Cheng:1993gf}.
  }
\begin{ruledtabular}    \begin{tabular}{cccc}
    &$\text{BR}(\Lambda_c\to\Lambda\pi^+)$&$\text{BR}(\Lambda_c\to\Sigma^0\pi^+$)&$\text{BR}(\Lambda_c\to\Sigma^+ \pi^0)$ \\
    \hline
PDG data~\cite{Tanabashi:2018oca}& $1.30\pm0.07 $& $1.29\pm0.07 $ & $1.24\pm0.10 $\\
BESIII~\cite{Ablikim:2015flg}& $1.24\pm0.07\pm0.03 $& $1.27\pm0.08\pm0.03 $ & $1.18\pm0.10\pm0.03$\\
SU(3)~\cite{Geng:2017esc}& $1.3\pm0.2 $& $1.3\pm0.2 $ & $1.3\pm0.2 $\\
Pole model~\cite{Cheng:1993gf}& $1.30\pm0.07 $& $1.29\pm0.07 $ & $1.24\pm0.10 $\\
Current algebra~\cite{Cheng:1993gf}& $1.30\pm0.07 $& $1.29\pm0.07 $ &$ 1.24\pm0.10 $\\
This work  & $1.30 $   & $1.24$  & $1.26 $\\
    \end{tabular}\end{ruledtabular}
     \label{tab:br}
    \end{center}
\end{table*}

The branching ratios of our final results are given in Table~\ref{tab:br}, where the PDG data~\cite{Tanabashi:2018oca}, BESIII new result~\cite{Ablikim:2015flg}, results based on the SU(3) flavor symmetry (SU(3))~\cite{Geng:2017esc}, pole model and current algebra~\cite{Cheng:1993gf} are also listed. It shows that the center values of our results are close to the experimental data within the conventional quark model parameter space.

\begin{table*}[ht]
  \begin{center}
  \renewcommand{\arraystretch}{1.3}
  \caption{ Uncertainties of the partial decay widths (in \%) caused by the quark model parameters with $20\%$ errors.}
\begin{ruledtabular}
\begin{tabular}{cccc}
 Input (GeV) &$\text{BR}(\Lambda_c\to\Lambda\pi^+)$
  &$\text{BR}(\Lambda_c\to\Sigma^0\pi^+)$
  &$\text{BR}(\Lambda_c\to\Sigma^+ \pi^0)$ \\
 \hline
$m_q=0.35\pm 0.070$    &$ 1.30\pm 0.46 $    &$ 1.24\pm 0.22 $  & $ 1.26\pm 0.23 $\\
$m_c=1.5\pm 0.30$      &$ 1.30\pm 0.011$    &$ 1.24\pm 0.053 $ & $1.26\pm 0.053 $\\
$\alpha'_\lambda=\alpha'_\rho=0.4\pm 0.08$  &$ 1.30\pm 0.50 $  &$ 1.24\pm 0.083 $  & $1.26\pm 0.082 $\\
$\alpha_\rho=0.45\pm 0.086$  &$ 1.30\pm 0.41$    &$ 1.24\pm 1.30 $  & $1.26\pm 1.32 $\\
$R=0.28\pm 0.056$     &$ 1.30\pm 1.01 $    &$ 1.24\pm 0.10 $   & $1.26\pm 0.10 $\\
Combined              &$ 1.30\pm 1.29 $    &$ 1.24\pm 1.33 $   &$  1.26\pm 1.35 $\\
    \end{tabular}\end{ruledtabular}
     \label{tab:err}
    \end{center}
\end{table*}

We also investigate the uncertainty sources by examining the sensitivities of the branching ratios to the model parameters which are listed in Table~\ref{tab:err}. The amplitude of direct pion emission is proportional to $R^{3/2}$ , while the dependence of $R$ for the CS amplitudes is more complicated and less sharp. A variation by  $20\%$ of the central value of $R$ leads to nearly $100\%$ change of the calculated branching ratio for $\Lambda_c\to \Lambda\pi^+$. Such a dramatic  sensitivity also indicates the dominance of the DPE process in $\Lambda_c\to \Lambda\pi^+$. In contrast, the impact of $R$ in $\Lambda_c\to \Sigma^0\pi^+$ and $\Sigma^+\pi^0$ turns out to be much less significant. This phenomenon is useful for examining the consistency of the model parameters since the experimental data can provide more stringent constraints on the model parameters.

One also notices the large uncertainties arising from the parameters $\alpha_\rho$ in the spatial wave function of the charmed baryons. It suggests that the branching ratios are more sensitive to the harmonic oscillator strengths than to the constituent quark masses. This is because of the strong dependence of the transition amplitudes on $\alpha_\rho$ in the wavefunction convolutions. Although the large uncertainties caused by $\alpha_\rho$ by varying 20\% of the adopted value may raise concerns about the quark model predictive power, this could also indicate that the hadronic weak decay observables are sensitive to the quark model parameters. Therefore, the hadronic weak decay processes may provide a better constraint on the quark model parameters. Further study of this interesting issue should be necessary to provide a more conclusive statement.

\begin{table*}[h]
  \begin{center}
  \renewcommand{\arraystretch}{1.3}
  \caption{ The asymmetry parameter $\alpha'$ and its uncertainties caused by the quark model parameters with $20\%$ errors.}
   \begin{ruledtabular} \begin{tabular}{cccc}
          &$\Lambda_c\to\Lambda\pi^+$
          &$\Lambda_c\to\Sigma^0\pi^+$
          &$\Lambda_c\to\Sigma^+ \pi^0$ \\
    \hline
     PDG data~\cite{Tanabashi:2018oca}  &$-0.91\pm 0.14$   &-              &$-0.45\pm0.32$\\
     Pole model~\cite{Cheng:1993gf}     &$-0.95$           &$0.78$         &$0.78$\\
     Current algebra~\cite{Cheng:1993gf}&$-0.99$           &$-0.49$        &$-0.49$\\
     This work                        &$-0.16\pm 0.27$   &$-0.46\pm0.20$ & $-0.47\pm0.19$\\
    \end{tabular}\end{ruledtabular}
     \label{tab:asy}
    \end{center}
\end{table*}

We can also calculate the parity asymmetry parameter in our model which is defined as
\begin{align}
\alpha'=\frac{2 \text{Re}(A^*B)}{|A|^2+|B|^2},
\end{align}
where $A$ and $B$ are the $S$ and $P$-wave amplitudes, respectively, defined at hadronic level. The hadronic level transition amplitude can be expressed as
\begin{eqnarray}
\label{eq:asy}
M'(B_i\to B_f + P)&=&i\bar u_f(m_f,\bm P_f)(A-B\gamma_5)u_i(m_i,\bm P_i) \nonumber\\
&\equiv &M'_{PV}(B_i\to B_f + P)+M'_{PC}(B_i\to B_f + P).
\end{eqnarray}
where the parity-violating and conserving amplitudes in the rest frame of the initial baryon can be written as,
\begin{align}
M'_{PV}(B_i\to B_f + P)&= i A\sqrt{\frac{E_f+m_f}{2m_f}} \chi^\dagger_f \chi_i ,\\
M'_{PC}(B_i\to B_f + P)&= i B\sqrt{\frac{E_f+m_f}{2m_f}} \chi^\dagger_f \frac{\bm \sigma \cdot \bm P_f}{E_f+m_f} \chi_i.
\end{align}
By comparing the above amplitudes with the corresponding quark model amplitudes we can determine $A$ and $B$. Then with the parity asymmetry parameter can be extracted:
\begin{align}
\alpha'=\frac{-2\text{Re}\left[(M'_{PV})^* M'_{PC}\right]}{|M'_{PC}|^2\dfrac{|\bm P_f|}{E_f+ m_f}+|M'_{PV}|^2 \dfrac{E_f+ m_f}{|\bm P_f|}}.
\end{align}
Namely, the amplitudes $M'_{PC/PV}$ can be expressed in terms of quark-model formalisms. The detailed expressions of $M'_{PC/PV}$ are given in Appendix~\ref{App:am}. In Tab.~\ref{tab:asy} the calculated parity asymmetries and uncertainties for these three channels are listed and compared with the PDG averaged values~\cite{Tanabashi:2018oca}, pole model calculation and current algebra treatment~\cite{Cheng:1993gf}. It shows that the result for $\Lambda_c\to \Sigma^+\pi^0$ agrees with the experimental data, while the value for the $\Lambda\pi^+$ appears to have quite significant discrepancies. Notice, however, that the $\Lambda\pi^+$ channel is sensitive to the DPE mechanism and the strong dependence of the pion wavefunction parameter $R$ can result in quite significant uncertainties. As a qualitative estimation we find that $\alpha'=-0.16\pm 0.27$ caused by the quark model parameters with $20\%$ and the error is larger than the other two channels. This, again, indicates the strong interfering effects between the DPE and non-factorizable amplitudes. In contrast, the uncertainties caused by $R$ in the $\Sigma\pi$ channels are much smaller due to the absence of the DPE process and relative suppression of the CS term relative to the pole terms.

\section{Summary}
\label{summary}

In this paper we investigate the two-body hadronic weak decay mechanism of $\Lambda_c$ in the framework of the non-relativistic constituent quark model. We first consider the Cabbibo-favored processes $\Lambda_c\to\Lambda\pi^+$, $\Sigma^0\pi^+$ and $\Sigma^+\pi^0$. These processes are correlated with each other and exhibit interesting features that can help disentangle the underlying dynamics. On the one hand, the $\Lambda\pi^+$ channel allows the DPE process which is factorizable and plays a dominant role, while the DPE process is absent in the $\Sigma\pi$ channels. On the other hand, these channels share some common features due to the SU(3) flavor symmetry in their non-factorizable transitions. With the availability of experimental data we find that the non-factorizable mechanisms from the pole terms and CS processes contribute the same order of magnitude as the DPE in $\Lambda_c\to\Lambda\pi^+$. This explains that the compatible branching ratios among these channels.

The coherent study of these processes is found useful for understanding the  structure of the baryons. In particular, we show that too strong a scalar-isoscalar $ud$ correlation in $\Lambda_c$ is not favored. Instead, it only needs to fulfill a quantum correlation in the spin-isospin and color space. Although the numerical results turn out to be sensitive to the parameters of the wave function parameters, a good understanding is reached based on the constituent quark effective degrees of freedom.

In the framework of the quark model, it is shown that there are destructive interferences between the A-type and B-type of pole terms in the transition amplitudes. This is similar to the case of light hyperon hadronic weak decays (e.g. see Ref.~\cite{Richard:2016hac} for the most recent detailed analysis of the $\Lambda$ and $\Sigma^\pm$ decays into nucleon and pion). Due to the destructive interferences it suggests that the SU(3) flavor symmetry breaking can become complicated. A relatively small symmetry breaking effects in each pole term can result in much more significant effects after the destructive interferences. This may explain why the current algebra treatment fails when describing some SU(3) flavor symmetry correlated channels~\cite{Cheng:1993gf}. Extension of this method to other hadronic weak decay channels may bring more insights into the role played by the non-factorizable processes in $\Lambda_c$ decays and provide more evidence for the quantum correlation for the light quarks. It is quite possible that other processes may provide a better constraint on the model uncertainties which will be investigated in the future.

\begin{acknowledgments}
We are grateful to Yu Lu for the help on the analytic calculation. We thanks H.-Y. Cheng for his interest in this work and useful feedbacks on an early version of this manuscript. This work is supported, in part, by the National Natural Science Foundation of China (Grant Nos. 11425525 and 11521505), DFG and NSFC funds to the Sino-German CRC 110 ``Symmetries and the Emergence of Structure in QCD'' (NSFC Grant No. 11261130311), Strategic Priority Research Program of Chinese Academy of Sciences (Grant No. XDB34030302), and National Key Basic Research Program of China under Contract No. 2015CB856700. J.M.R. would like to thank the hospitality provided to him at IHEP, where part of this work was completed, and the support  by the Munich Institute for Astro- and Particle Physics (MIAPP) of the DFG cluster of excellence ``Origin and Structure of the Universe'' during the Workshop ``Deciphering Strong-Interaction Phenomenology through Precision Hadron-Spectroscopy.''
Q.W. is also supported by the research startup funding at SCNU, Guangdong Provincial funding with Grant No. 2019QN01X172£¬and Science and Technology Program of Guangzhou (No. 2019050001).
\end{acknowledgments}

\begin{appendix}
\section{The harmonic oscillator Hamiltonian and the Jacobi coordinates}
\label{app:JAC}
Here, we briefly summarize our notations for the Jacobi coordinates that are use to separate the center-of-mass motion in non-relativistic models, and treat explicitly the harmonic-oscillator model that is used to parameterize the baryon wave functions. There are several variants. Let us first follow~\cite{Isgur:1978xj}. The Hamiltonian is
\begin{equation}
\label{eq:Ham}
H=\sum_{i=1}^3 \frac{\bm p_i^2}{2 m_i}+\frac{1}{2}K \sum_{i<j} (\bm r_i-\bm r_j)^2,
\end{equation}
where $\bm p_i$, $\bm r_i$ and $m_i$ denote the momentum, position and mass of the $i$-th quark,  and $K$ is the the spring constant.  With $m_1=m_2=m$ and $m_3=m'$, the  Jacobi coordinates are defined as~\cite{Nagahiro:2016nsx}:
\begin{equation}
\label{eq:transition}
\left\{
\begin{aligned}
\bm R_c &= \frac{1}{M}\left( m \bm r_1+m \bm r_2+m' \bm r_3 \right) \\
\bm \rho &=\bm r_1 -\bm r_2 \\
\bm \lambda &= \frac{1}{2} \left( \bm r_1 +\bm r_2 -2\bm r_3\right)
\end{aligned}\right.
\qquad\qquad
\left\{
\begin{aligned}
\bm P&= \bm p_1+\bm p_2+\bm p_3\\
\bm p_\rho&= \frac{1}{2} \left( \bm p_1- \bm p_2 \right)\\
\bm p_\lambda&=\frac{1}{M} \left(m' \bm p_1+m' \bm p_2-2m\bm p_3\right)
\end{aligned}\right. \ .
\end{equation}
and the Hamiltonian becomes
\begin{align}
\label{eq:HamJ}
H=\frac{\bm P^2}{2M}+\frac{\bm p_\rho^2}{2 m_\rho}+\frac{\bm p_\lambda^2}{2 m_\lambda}+\frac{1}{2} m_\rho \omega_\rho^2 \bm \rho^2 +\frac{1}{2} m_\lambda \omega_\lambda^2 \bm \lambda^2,
\end{align}
where $M=m_1+m_2+m_3$, $m_\rho=m/2$ and $m_\lambda= 2m m'/M$ are the reduced masses of the $\rho$ and $\lambda$ degrees of freedom, respectively; $\omega_\rho= \sqrt{3K/m}$ and $\omega_\lambda=\sqrt{2K/m_\lambda}$ are the frequencies of the corresponding harmonic oscillators.

Then, the spatial wave functions on the harmonic oscillator basis can be obtained~\cite{LeYaouanc:1988fx,Pervin:2005ve,Zhong:2007gp,Nagahiro:2016nsx}. In the coordinate space, a basis for the eigen wavefunctions is:
\begin{align}
\Psi_{N,L,L_z}(\bm R_c,\bm \rho,\bm\lambda)
=\frac{1}{(2\pi)^{3/2}}
\exp\left(-i \bm P\cdot \bm R_c\right)
\sum_m \langle l_\rho,m;l_\lambda,L_z-m|L,L_z \rangle\tilde\psi^{\alpha_\rho}_{n_\rho l_\rho m }(\bm \rho)\tilde\psi^{\alpha_\lambda}_{n_\lambda l_\lambda L_z-m }(\bm \lambda),
\end{align}
where $N$ stands for $\{n_\rho,l_\rho;n_\lambda,l_\lambda\}$, and
\begin{align}
\tilde\psi^\alpha_{nlm}(\bm r)=\left[\frac{2\,n!}{(n+l+1/2)!} \right]^{1/2}\alpha^{l+3/2}
\exp\left(-\frac{\alpha^2\bm r^2}{2}\right)L_n^{l+1/2}(\alpha^2\bm r^2)
\mathcal{Y}_{lm}(\bm r),
\end{align}
where $\bm P$ is the total momentum of the three quark system. The function $L_n^\nu(x)$ is the generalized Laguerre polynomial, and $\alpha_\rho$ and $\alpha_\lambda$ are the harmonic oscillator strengths defined by
\begin{align}
\label{eq:hos}
\alpha_\rho^2=m_\rho \omega_\rho=\frac{\sqrt{3K m}}{2}, \ \ \alpha_\lambda^2=m_\lambda \omega_\lambda=2\sqrt{K \frac{m m'}{M}} \ .
\end{align}

In the momentum space the spatial wave function is written as:
\begin{align}
\Psi_{N L L_z}(\bm P,\bm p_\rho,\bm p_\lambda)
=\delta^3(\bm P-\bm P_c)\sum_m \langle l_\rho,m;l_\lambda,L_z-m|L,L_z \rangle
\psi^{\alpha_\rho}_{n_\rho l_\rho m }(\bm p_\rho)
\psi^{\alpha_\lambda}_{n_\lambda l_\lambda L_z-m }(\bm p_\lambda),
\end{align}
where
\begin{align}
\psi^\alpha_{nlm}(\bm p)=(i)^l(-1)^n \left[\frac{2n!}{(n+l+1/2)!} \right]^{1/2}\frac{1}{\alpha^{l+3/2}}\exp\left({-\frac{\bm p^2}{2\alpha^2}}\right)L_n^{l+1/2}(\bm p^2/\alpha^2)
\mathcal{Y}_{lm}(\bm p).
\end{align}

One can also choose a slightly differently scaled Jacobi coordinates which are more convenient to implement the permutation properties. $R_c$ and $P$ are identical, but now
\begin{equation}\label{Jacobi-coordinate-2}
\left\{
\begin{aligned}
\tilde{\bm \rho} &=\frac{1}{\sqrt 2}(\bm r_1 -\bm r_2)\\
\tilde{\bm \lambda} &=\frac{1}{\sqrt 6}(\bm r_1 + \bm r_2- 2\bm r_3)
\end{aligned}\right.,\qquad\qquad
 \left\{
 \begin{aligned}
 \bm p_\rho&= \frac{1}{\sqrt 2} (\bm p_1-\bm p_2)\\
 \bm p_\lambda&=\frac{3}{\sqrt6 M}( m' \bm p_1+ m' \bm p_2-2m\bm p_3)
 \end{aligned},\right.\ ,
\end{equation}
The  reduced masses are now $\tilde m_\rho=m$ and $\tilde m_\lambda=3m m'/M$. The frequencies and oscillator strengths become
\begin{equation}
\begin{alignedat}{2}
\tilde\omega_\rho &=\sqrt{\frac{3K}{\tilde m_\rho}},\qquad&&\tilde\omega_\lambda=\sqrt{\frac{3K}{\tilde m_\lambda}},
\\
\tilde\alpha_\rho^2&=\sqrt{3K m}, && \tilde\alpha_\lambda^2=3\sqrt{K \frac{m m'}{M}},
\end{alignedat}
\end{equation}
the correspondence being
\begin{equation}
\alpha_\rho= \frac{\tilde\alpha_\rho}{\sqrt2}, \ \ \ \alpha_\lambda=\sqrt{\frac{2}{3}}\tilde\alpha_\lambda.
\end{equation}

\section{Wave functions}
\label{App:wavefuncion}
In the framework of the non-relativistic constituent quark model, the wave functions of baryons or mesons consist of four parts: (i) color; (ii) flavor; (iii) spin, and (iv) spatial wave function. The color wave function is unique for non-exotic color-singlet hadrons. We only list the spin, flavor and spatial wave functions. In the light sector, it is useful to identify the  behavior with respect to the permutation group $s_3$.

\subsection{Baryon wave functions}

The spin wave functions for baryons are:
\begin{equation}\begin{aligned}
\chi^{\rho}_{\frac{1}{2},\frac{1}{2}}&=\frac{1}{\sqrt2}\left(\uparrow \downarrow \uparrow -\downarrow \uparrow \uparrow \right) \ ,
\quad& \chi^{\lambda}_{\frac{1}{2},\frac{1}{2}}&=-\frac{1}{\sqrt6}\left( \uparrow\downarrow\uparrow+\downarrow\uparrow\uparrow-2\uparrow\uparrow\downarrow \right) \ , \\
\chi^{\rho}_{\frac{1}{2},-\frac{1}{2}}&=\frac{1}{\sqrt2}\left(\uparrow \downarrow \downarrow -\downarrow \uparrow \downarrow \right) \ ,
&\chi^{\lambda}_{\frac{1}{2},-\frac{1}{2}}&=\frac{1}{\sqrt6}\left( \uparrow\downarrow\downarrow+\downarrow\uparrow\downarrow-2\downarrow\downarrow\uparrow\right) \ .
\end{aligned}\end{equation}
\begin{equation}\begin{aligned}
\chi^{s}_{\frac{3}{2},\frac{3}{2}}&=\uparrow \uparrow \uparrow \ ,
\quad& \chi^{s}_{\frac{3}{2},-\frac{3}{2}}&= \downarrow\downarrow\downarrow \ , \\
\chi^{s}_{\frac{3}{2},\frac{1}{2}}&=\frac{1}{\sqrt3}\left(\uparrow \uparrow \downarrow +\uparrow \downarrow \uparrow +\downarrow \uparrow \uparrow \right) \ ,
&\chi^{s}_{\frac{3}{2},-\frac{1}{2}}&=\frac{1}{\sqrt3}\left( \uparrow\downarrow\downarrow+\downarrow\uparrow\downarrow+\downarrow\downarrow\uparrow\right) \ .
\end{aligned}\end{equation}
The symbol $\rho$ and $\lambda$ are used to label the two components of the mixed-symmetry pair. The symbol $s$ is used to label the symmetric states.

The flavor wave functions for $\Lambda$, $\Sigma^0$ and $\Sigma^+$ as the SU(3) flavor octet states~\cite{LeYaouanc:1988fx} are:
\begin{equation}\begin{aligned}
\phi_{\Lambda}^\lambda&=-\frac{1}{2}(sud+usd-sdu-dsu) \ ,
\quad&\phi_{\Lambda}^\rho&=\frac{1}{2\sqrt3}(usd+sdu-sud-dsu-2dus+2uds) \ , \\
\phi_{\Sigma^+}^\lambda &= \frac{1}{\sqrt6}(2uus-suu-usu) \ ,
&\phi_{\Sigma^+}^\rho &=\frac{1}{\sqrt2}(suu-usu) \ ,\\
\phi_{\Sigma^0}^\lambda&=\frac{1}{2\sqrt3}(sdu+sud+usd+dsu-2uds-2dus) \ ,
&\phi_{\Sigma^0}^\rho&=\frac{1}{2}(sud+sdu-usd-dsu)\ .
\end{aligned}\end{equation}

For the flavor wave functions of charmed baryons there are two bases adopted in the literature. One is the ``$uds$'' basis which is used in our calculation. Namely, similar to the hyperon wave functions, the flavor wave functions of $\Lambda_c$ and $\Sigma_c^0$ are obtained by making the replacement of $s\rightarrow c $ in the above  hyperon wave functions~\cite{Hussain:1983pk}.

The other one is the ``$udc$'' basis~\cite{Copley:1979wj,Zhong:2007gp} in which  only the symmetry among the light quarks is implemented. It reads
\begin{equation}
\phi_{\Lambda_c}=\frac{1}{\sqrt2}(ud-du)c \ , \qquad
\phi_{\Sigma_c} =\begin{dcases}
ddc & \text{for } \Sigma_c^0 \ , \\
\frac{1}{\sqrt2}(ud+du)c & \text{for } \Sigma_c^+ \ ,\\
uuc & \text{for } \Sigma_c^{++} \ .
\end{dcases}
\end{equation}

With the spin, flavor and spatial parts,  we can construct the total wave function of the baryons, which is denoted $|B \prescript{2S+1}{}{L}J^P \rangle$. In the light sector, the ground state reads
\begin{equation}
|B \prescript{2}{}{S}\half^+ \rangle=\frac{1}{\sqrt2}(\phi_B^\rho \chi^\rho_{S,S_z}+\phi_B^\lambda \chi^\lambda_{S,S_z})\Psi_{0,0,0} \ ,
\label{eq:og}
\end{equation}
and for charmed baryons
\begin{equation}\begin{aligned}
|\Lambda_c ^{2}{S}\half^+ \rangle &={} \phi_{\Lambda_c} \chi^\rho_{S,S_z}\Psi_{0,0,0} \ ,\\
|\Sigma_c ^{2}{S}\half^+ \rangle &={} \phi_{\Sigma_c} \chi^\lambda_{S,S_z}\Psi_{0,0,0}\ .
\end{aligned}\end{equation}

 For the first orbital excitation states, we have two different modes, i.e., $\rho$ and  $\lambda$ configurations. In the light sector, they are recombined into the single symmetric state
\begin{align}
|B \prescript{2}{}{P}\half^- \rangle &=\sum_{L_z+S_z=J_z}\langle1, L_z;\half, S_z|J J_z\rangle\frac{1}{2}\left[(\phi_B^\rho \chi_{S,S_z}^\lambda+\phi_B^\lambda \chi_{S,S_z}^\rho)\Psi^\rho_{1,L_z}+(\phi_B^\rho \chi_{S,S_z}^\rho-\phi_B^\lambda \chi_{S,S_z}^\lambda)\Psi^\lambda_{1,L_z} \right] , \\
|B \prescript{3}{}{P}\half^- \rangle &=\sum_{L_z+S_z=J_z}\langle1, L_z;\frac{3}{2}, S_z|J J_z\rangle\frac{1}{\sqrt 2}\left[\phi_B^\rho \chi_{S,S_z}^s \Psi^\rho_{1,L_z}+ \phi_B^\lambda \chi_{S,S_z}^s\Psi^\lambda_{1,L_z} \right] ,
\label{eq:of}
\end{align}
where $\Psi^\lambda_{1,L_z}$ stands for $\Psi_{N,L_z}$ with $N=\{0,0;0,1\}$ and $\Psi^\rho_{1,L_z}$ corresponds to $N=\{0,1;0,0\}$.
In the charm sector, they read
\begin{equation}
\begin{aligned}|\Lambda_c ^{2}{P}_\lambda\half^-\rangle &=& \sum_{L_z+S_z=J_z}\la 1, L_z;\half, S_z \middle|\half, J_z \ra \phi_{\Lambda_c} \chi^\rho_{S,S_z} \Psi^{\lambda}_{1,L_z} \ , \\
|\Lambda_c ^{2}{P}_\rho\half^-\rangle &=& \sum_{L_z+S_z=J_z}\la 1, L_z;\half S_z \middle|\half, J_z \ra \phi_{\Lambda_c} \chi^\lambda_{S,S_z} \Psi^{\rho}_{1,L_z} \ , \\
|\Sigma_c ^{2}{P}_\lambda\half^-\rangle &=& \sum_{L_z+S_z=J_z}\la 1, L_z;\half, S_z \middle|\half, J_z \ra \phi_{\Sigma_c} \chi^\lambda_{S,S_z} \Psi^{\lambda}_{1,L_z} \ , \\
|\Sigma_c ^{2}{P}_\rho\half^-\rangle &=& \sum_{L_z+S_z=J_z}\la 1, L_z;\half, S_z \middle|\half, J_z \ra \phi_{\Sigma_c} \chi^\rho_{S,S_z} \Psi^{\rho}_{1,L_z} \ .
\end{aligned}\end{equation}

\subsection{Pion wave function}
The wave function of pseudoscalar mesons is written as:
\begin{align}
\Phi_{0,0,0} (\bm p_1,\bm p_2)=\delta^3(\bm p_1+\bm p_2-\bm P)\phi_p \chi^a_{0,0}\psi_{0,0,0}(\bm p_1,\bm p_2),
\end{align}
where $\chi^a_{0,0}$ is the spin wave:
\begin{align}
\chi^a_{0,0}=\frac{1}{\sqrt2}(\uparrow \downarrow-\downarrow \uparrow),
\end{align}
and $\phi_p (p=\pi^+,\pi^0,\pi^-)$ is the flavor wave function
\begin{align}
\phi_{\pi^+}&=u \bar d, \\
\phi_{\pi^0}&=-\frac{1}{\sqrt2}(u \bar u - d \bar d),\\
\phi_{\pi^-}&=-d\bar u .
\end{align}
The spatial wave function is expressed as:
\begin{equation}
\psi_{0,0,0}(\bm p_1, \bm p_2)= \frac{1}{\pi^{3/4} R^{3/2}}\exp\left[-\frac{(\bm p_1-\bm p_2)^2}{8 R^2}\right],
\end{equation}
where $R$ is the parameter of the meson wave function.
\section{Amplitudes for the pole terms}\label{App:am}
The transition amplitudes denoted by the baryon polarization quantum numbers are to be provided. In addition, $M^{J_f,J^z_f;J_i,J^z_i}$ is shortened to $M^{J^z_f;J^z_i}$ as the spin of initial and final states are all $1/2$. We will provide the expressions for the amplitude of $M^{-1/2,-1/2}$ for each process and the Hermitian relation gives: $M_{PC}^{1/2,1/2}=-M_{PC}^{-1/2,-1/2}$ and $M_{PV}^{1/2,1/2}=M_{PV}^{-1/2,-1/2}$. The amplitudes of $M^{\pm1/2,\mp1/2}$ are vanishing. It should be noted that the pole term processes are two vertex process while the CS and DPE processes are one vertex process. So the relative phase difference between these two types processes is $\pi$.

In the results given below, we use the second set of Jacobi coordinates of Appendix~\ref{app:JAC}, but to alleviate the writing, the tildes are omitted for the $\alpha$'s. The following functions are to be used later:
\begin{align}
&\xi=\left(\frac{4 \alpha_\lambda \alpha_\rho}{4\alpha^2+\alpha_\lambda^2+3\alpha^2_\rho}\right)^{3/2},
&&\mathcal F_{\pi}(k)=\exp\left[-\frac{k^2}{6 \alpha^2}\right],
&&&\mathcal F_{\pi}'(k)=\exp\left[-\frac{k^2}{24}\left(\frac{1}{\alpha_\lambda^2}+\frac{3}{\alpha_\rho^2}\right)\right],
\end{align}
where $m_q$ is the mass of the light quarks ($u,d,s$) and $m_c$ is the mass of the $c$ quark; $k\equiv |\bm k|$ and $\omega_0$ denote the three-vector momentum and energy of the pion, respectively. In order to use the typical value of the harmonic oscillator strengths directly, all the amplitudes are expressed with the conventional the harmonic oscillator strengths. $\alpha_\rho$ and $\alpha_\lambda$ are the harmonic oscillator strengths for the charmed baryons and $\alpha=\alpha_\lambda'=\alpha_\rho'$ for the light baryons. For the pole terms, the propagator is noted with $\mathcal P(m_1,m_2)$ which is defined as
\begin{align}
\mathcal P (m_1,m_2)=\frac{ 2m_2}{m_1^2-m_2^2+i \Gamma_{m_2} m_2},
\end{align}
where $m_1$ is the mass of initial baryon or final baryon and $m_2$ is the mass of intermediate baryons. $\Gamma_{m_2}$ is the width of intermediate baryons.

\subsection{$\Lambda_c\to \Lambda \pi^+$}
a) pole terms
\begin{align}
M_{\text{Pole,A};PC}^{-1/2;-1/2}=&
\left[\sqrt3 V_{ud}V_{cs}G_F\frac{\alpha^3}{\pi^{3/2}} \xi\right]
\left[- \frac{k(6m_q+\omega_0)}{12 \sqrt 6\pi^{3/2}\sqrt{\omega_0}f_\pi m_q} \mathcal F_{\pi}(k) \right]
\mathcal P(m_{\Lambda_c},m_{\Sigma^+}), \\
M_{\text{Pole,A1};PV}^{-1/2;-1/2}=&
\left[i\sqrt{\frac23}V_{ud}V_{cs}G_F \frac{\alpha^4}{\pi^{3/2}}\frac{(-6\alpha^2+\alpha_\lambda^2-15\alpha_\rho^2)m_c+(-6\alpha^2+5\alpha_\lambda^2-3\alpha_\rho^2)m_q}{2 m_c m_q(4\alpha^2+\alpha_\lambda^2+3\alpha_\rho^2)} \xi \right]
\notag \\
&\times\left[-i\frac{k^2(6m_q+\omega_0)-18\alpha^2\omega_0}{144\sqrt3 \pi^{3/2}\sqrt{\omega_0}f_\pi m_q \alpha}\mathcal F_{\pi}(k) \right]
\mathcal P(m_{\Lambda_c},m_{\Sigma^{*+}}), \\
M_{\text{Pole,A2};PV}^{-1/2;-1/2}=&
\left[-i\sqrt{\frac23}V_{ud}V_{cs}G_F \frac{\alpha^4}{\pi^{3/2}}\frac{(6\alpha^2+\alpha_\lambda^2+3\alpha_\rho^2)m_c+2(3\alpha^2+\alpha_\lambda^2+3\alpha_\rho^2)m_q}{ m_c m_q(4\alpha^2+\alpha_\lambda^2+3\alpha_\rho^2)} \xi \right]
\notag \\
&\times\left[i\frac{k^2(6m_q+\omega_0)-18\alpha^2\omega_0}{72\sqrt3 \pi^{3/2}\sqrt{\omega_0}f_\pi m_q \alpha}\mathcal F_{\pi}(k) \right]
\mathcal P(m_{\Lambda_c},m_{\Sigma(1750)}), \\
M_{\text{Pole,B};PC}^{-1/2;-1/2}=&
\left[\frac{k(6m_q +\omega_0)}{18\sqrt6 \pi^{3/2} \sqrt{\omega_0}f_\pi m_q}\mathcal F'_{\pi}(k) \right]
\left[-\sqrt 3 V_{ud}V_{cs}G_F \frac{\alpha^3}{\pi^{3/2}}~\xi   \right]
\mathcal P(m_{\Lambda},m_{\Sigma_c^0}), \\
M_{\text{Pole,B};PV}^{-1/2;-1/2}=
&\left[i\frac{18\alpha_\rho\alpha_\lambda(\alpha_\lambda+3\alpha_\rho)\omega_0+(\alpha_\rho+3\alpha_\lambda)k^2(6m_q+\omega_0)}{864\sqrt3\pi^{3/2}\alpha_\lambda\alpha_\rho f_\pi \sqrt{\omega_0}m_q}\mathcal F'_{\pi}(k) \right]\notag \\
&\times\left[-i V_{cs}V_{ud}G_F\frac{\sqrt 6}{\pi^{3/2} m_q}\frac{\alpha^3(\alpha_\rho+\alpha_\lambda)(\alpha^2+\alpha_\rho\alpha_\lambda)}{4\alpha^2+\alpha_\lambda^2+3\alpha^2_\rho} ~\xi\right]
\mathcal P(m_{\Lambda},m_{\Sigma_C^{*0}}).
\end{align}

b) direct pion emission term
\begin{align}
M_{DPE,PC}^{-1/2;-1/2}
&=-\frac{\sqrt2}{3}V_{ud}V_{cs}G_F\frac{k}{\pi^{9/4}m_q}\frac{3\alpha^2+5\alpha_\lambda^2}{\alpha^2+\alpha_\lambda^2}\left[\frac{\alpha^2 \alpha_\lambda \alpha_\rho R}{(\alpha^2+\alpha_\lambda^2)(\alpha^2+\alpha^2_\rho)}\right]^{3/2} \exp\left[-\frac{k^2}{3(\alpha^2+\alpha_\lambda^2)}\right],
 \\ \notag\\
M_{DPE,PV}^{-1/2;-1/2}
&=2 \sqrt2 \frac{V_{ud}V_{cs}G_F}{\pi^{9/4}}\left[\frac{\alpha^2 \alpha_\lambda \alpha_\rho R}{(\alpha^2+\alpha_\lambda^2)(\alpha^2+\alpha^2_\rho)}\right]^{3/2} \exp\left[-\frac{k^2}{3(\alpha^2+\alpha_\lambda^2)}\right] .
\end{align}

c) color suppressed terms
\begin{align}
M_{CS,PC}^{-1/2;-1/2}&=2\sqrt3 V_{cs}V_{ud}G_F k \left(\alpha^2 \alpha_\lambda \alpha_\rho R\right)^{3/2} \notag \\
&\times \frac{m_c\left( \alpha^2(\alpha^2_\lambda+3\alpha^2_\rho)+
3\alpha^2_\lambda\alpha^2_\rho+2R^2(6\alpha^2+2\alpha_\lambda^2+3\alpha^2_\rho)+m_q\alpha_\lambda^2(2\alpha^2+\alpha_\rho^2-2R^2) \right)}{\pi^{9/4} m_c m_q\left[ 2\alpha^2(\alpha^2_\lambda+3\alpha^2_\rho)+
6\alpha^2_\lambda\alpha^2_\rho+3R^2(4\alpha^2+\alpha_\lambda^2+3\alpha^2_\rho) \right]^{5/2}} \notag \\
&\times\exp\left[-\frac{k^2}{24}\frac{36\alpha^2+25\alpha^2_\lambda+3\alpha^2_\rho+24R^2}{2\alpha^2(\alpha^2_\lambda+3\alpha^2_\rho)+
6\alpha^2_\lambda\alpha^2_\rho+3R^2(4\alpha^2+\alpha_\lambda^2+3\alpha^2_\rho)}\right], \\
\notag \\
M_{CS,PV}^{-1/2;-1/2}&=-4 \sqrt 3V_{cs}V_{ud}G_F
\left[\frac{\alpha^2 \alpha_\lambda \alpha_\rho R}{2\alpha^2(\alpha^2_\lambda+3\alpha^2_\rho)+
6\alpha^2_\lambda\alpha^2_\rho+3R^2(4\alpha^2+\alpha_\lambda^2+3\alpha^2_\rho)}\right]^{3/2} \notag \\
&\times \exp\left[-\frac{k^2}{24}\frac{36\alpha^2+25\alpha^2_\lambda+3\alpha^2_\rho+24R^2}{2\alpha^2(\alpha^2_\lambda+3\alpha^2_\rho)+
6\alpha^2_\lambda\alpha^2_\rho+3R^2(4\alpha^2+\alpha_\lambda^2+3\alpha^2_\rho)}\right].
\end{align}

\subsection{$\Lambda_c\to \Sigma^0 \pi^+$ and $\Lambda_c\to \Sigma^0 \pi^+$ }
The amplitudes of $\Lambda_c\to \Sigma^0 \pi^+$ and $\Lambda_c\to \Sigma^0 \pi^+$ have the same form. In the following, only the amplitudes of $\Lambda_c\to \Sigma^0 \pi^+$ are given.

a) pole terms
\begin{align}
M_{\text{Pole,A};PC}^{-1/2;-1/2}=&
\left[\sqrt3 V_{ud}V_{cs}G_F\frac{\alpha^3}{\pi^{3/2}} \xi\right]
\left[\frac{k(6m_q+\omega_0)}{18 \sqrt 2\pi^{3/2}\sqrt{\omega_0}f_\pi m_q} \mathcal F_{\pi}(k) \right]
\mathcal P(m_{\Lambda_c},m_{\Sigma^+}), \\
M_{\text{Pole,A1},PV}^{-1/2;-1/2}=&
\left[i\sqrt{\frac23}V_{ud}V_{cs}G_F \frac{\alpha^4}{\pi^{3/2}}\frac{(-6\alpha^2+\alpha_\lambda^2-15\alpha_\rho^2)m_c+(-6\alpha^2+5\alpha_\lambda^2-3\alpha_\rho^2)m_q}{2 m_c m_q(4\alpha^2+\alpha_\lambda^2+3\alpha_\rho^2)} \xi \right]
\notag \\
&\times \left[i \frac{5k^2(6m_q+\omega_0)-18\alpha^2\omega_0}{432 \pi^{3/2}\sqrt{\omega_0}f_\pi m_q \alpha}\mathcal F_{\pi}(k) \right]
\mathcal P(m_{\Lambda_c},m_{\Sigma^{*+}}), \\
M_{\text{Pole,A2};PV}^{-1/2;-1/2}=&
\left[-i\sqrt{\frac23}V_{ud}V_{cs}G_F \frac{\alpha^4}{\pi^{3/2}}\frac{(6\alpha^2+\alpha_\lambda^2+3\alpha_\rho^2)m_c+2(3\alpha^2+\alpha_\lambda^2+3\alpha_\rho^2)m_q}{ m_c m_q(4\alpha^2+\alpha_\lambda^2+3\alpha_\rho^2)} \xi \right]
\notag \\
&\times\left[i\frac{k^2(6m_q+\omega_0)-18\alpha^2\omega_0}{216 \pi^{3/2}\sqrt{\omega_0}f_\pi m_q \alpha}\mathcal F_{\pi}(k) \right]
\mathcal P(m_{\Lambda_c},m_{\Sigma(1750)}) ,\\
M_{\text{Pole,B};PC}^{-1/2;-1/2}=&
\left[\frac{k(6m_q +\omega_0)}{18\sqrt6 \pi^{3/2} \sqrt{\omega_0}f_\pi m_q}\mathcal F'_{\pi}(k) \right]
\left[3 V_{ud}V_{cs}G_F \frac{\alpha^3}{\pi^{3/2}}~\xi   \right]
\mathcal P(m_{\Sigma^0},m_{\Sigma_c^0}), \\
M_{\text{Pole,B};PV}^{-1/2;-1/2}=
&\left[i\frac{18\alpha_\rho\alpha_\lambda(\alpha_\lambda+3\alpha_\rho)\omega_0+(\alpha_\rho+3\alpha_\lambda)k^2(6m_q+\omega_0)}{864\sqrt3\pi^{3/2}\alpha_\lambda\alpha_\rho f_\pi \sqrt{\omega_0}m_q} F'_{\pi}(k) \right]\notag \\
&\times \left[i V_{cs}V_{ud}G_F\frac{3\sqrt 2}{\pi^{3/2} m_q}\frac{\alpha^3(\alpha_\rho+\alpha_\lambda)(\alpha^2+\alpha_\rho\alpha_\lambda)}{4\alpha^2+\alpha_\lambda^2+3\alpha^2_\rho} ~\xi\right]
\mathcal P(m_{\Sigma^0},m_{\Sigma_c^{*0}}).
\end{align}

b) color suppressed term
\begin{align}
M_{CS,PC}^{-1/2;-1/2}(\Lambda_c\to \Sigma^0 \pi^+)=-\sqrt 3 M_{CS,PC}^{-1/2;-1/2}(\Lambda_c\to \Lambda \pi^+), \\
M_{CS,PV}^{-1/2;-1/2}(\Lambda_c\to \Sigma^0 \pi^+)=-\sqrt 3 M_{CS,PV}^{-1/2;-1/2}(\Lambda_c\to \Lambda \pi^+).
\end{align}

\end{appendix}

\end{document}